\documentclass[fleqn,11pt]{article}

\usepackage{amsmath,amsfonts,amssymb}
\newcommand\numberthis{\addtocounter{equation}{1}\tag{\theequation}}
\usepackage{color}

\usepackage{geometry}
\newtheorem{theorem}{Theorem}
\newtheorem{lemma}[theorem]{Lemma}

\newtheorem{proposition}{Proposition}

\renewcommand{\theequation}{\thesection.\arabic{equation}}

\newcommand{\R}{\mathbb{R}}

\newcommand{\N}{\mathbb{N}}
\newcommand{\T}{\mathbb{T}}

\newcommand{\cqfd}
{%
\mbox{}%
\nolinebreak%
\hfill%
\rule{2mm}{2mm}%
\newline
\newline
}

\title{Bose condensates in interaction with excitations - a two-component space-dependent model close to equilibrium.}
\author{Leif ARKERYD and Anne NOURI\\
\\Mathematical Sciences, 41296 G\"oteborg, Sweden,\\
arkeryd@chalmers.se\\
Aix-Marseille University, CNRS, Centrale Marseille, I2M UMR 7373, 13453 Marseille, France,\\
anne.nouri@univ-amu.fr}

\date{}

\begin{document}

\maketitle

{\noindent \bf Abstract.}\hspace{0.1in}
 The paper considers a model for Bose gases in the so-called 'high-temperature range' below the temperature where Bose-Einstein condensation sets in. The model is of non-linear two-component type, consisting of a kinetic equation with periodic boundary conditions for the distribution function of a gas of excitations interacting with a Bose condensate, which is described by a Gross-Pitaevskii equation. Results on well-posedness and long time behaviour are proved in a Sobolev space setting close to equilibrium.

\footnotetext[2]{Key words; low temperature kinetics, Bose condensate, two component model, 0.7 $T_c$.}

%
%
%
%
%
%
%
\section{ Preliminaries and main results.}
\subsection{Physics motivations.} The phenomenon of Bose-Einstein condensation occurs when a large number of particles of a Bose gas enter the same lowest accessible quantum state. Predicted by Bose and Einstein in 1924 [\ref{[B]}] [\ref{[E]}], it was first unambiguously produced in 1995 by E. Cornell and C. Wieman.
This paper studies a Bose condensate below the transition temperature $T_c$ for condensation, and in interaction with a non-condensates component. The setting is a two-component space-dependent model well established in physics (see the monograph [\ref{[GNZ]}] and its references) of pair-collision interactions involving a gas of thermally excited (quasi-)particles and a condensate. The two-component model consists of a kinetic equation for the distribution function of the gas, and a Gross-Pitaevskii equation (cf [\ref{[PS]}]) for the condensate.
A rather general form of the kinetic equation in the superfluid frame is (cf [\ref{[PBMR]}], [\ref{[ZNG]}])
\begin{eqnarray}
\partial _tf+ (\nabla_p(E(p))+v_c)\cdot\nabla_x f-\nabla_x(E(p)+v_c\cdot p)\cdot\nabla_pf = C_{22}(f)+C_{12}(f,n_c).
\end{eqnarray}
Here $f$ is the quasi-particle phase space density, $n_c$ (resp. $v_c$) is the mass density (resp. the velocity) of the condensate, and $E(p)$ denotes the (Bogoliubov) quasi-particle energy.
The Nordheim-Uehling-Uhlenbeck term $C_{22}$ for collisions between (quasi-)particles is given by
\begin{eqnarray}
{C}_{22}(f)(p)= \frac{g^2}{\hbar }\int _{\R ^3\times \R ^3\times \R ^3}B\delta (p+p_*= p'+p'_*)\delta (E(p)+E(p_*)= E(p')+E(p'_*))\nonumber \\
\Big( f'f'_*(1+f)(1+f_*)-ff_*(1+f')(1+f'_*)\Big) dp_*dp'dp'_*,
\end{eqnarray}
where the interaction strength $g= \frac{4\pi a\hbar ^2}{m}$, $a$ is the scattering length of the interaction potential, $\hbar $ the Planck constant, $m$ the atomic mass, $B$ a collision kernel, and
\begin{eqnarray*}
f= f(p),\quad f_*= f(p_*),\quad f'= f(p'),\quad f'_*= f(p'_*).
\end{eqnarray*}
The collision term $C_{12}$ for collisions between (quasi-)particles and condensate is
\begin{eqnarray}
C_{12}(f,n_c)(p)=
\frac{g^2n_c}{\hbar }\int_{\R ^3\times \R ^3\times \R ^3} A\delta(p_1-p_2-p_3)\delta(E_1-E_2-E_3)[\delta(p-p_1)\\
-\delta(p-p_2)-\delta(p-p_3)]
((1+f_1)f_2f_3-f_1(1+f_2)(1+f_3))dp_1dp_2dp_3,\nonumber
\end{eqnarray}
where $A$ is a collision kernel and
\begin{eqnarray*}
f_j=f(p_j),\quad E_j= E(p_j),\quad 1\leq j\leq 3.
\end{eqnarray*}
The usual Gross-Pitaevskii equation for the wave function $\psi$ (the order parameter) associated with a Bose condensate is
\begin{eqnarray*}
i\hbar \partial _t\psi = -\frac{\hbar^2}{2m}\Delta_x\psi+(g|\psi|^2+U_{ext})\psi,
\end{eqnarray*}
where $U_{ext}$ is an external potential, i.e. a Schr\"odinger equation complemented by a non-linear term accounting for two-body interactions.\\
In the present context, the Gross-Pitaevskii equation is further generalized by letting the condensate move in a self-consistent Hartree-Fock mean field $2\int_{\R ^3}\ f(p)dp$ produced by the thermally excited atoms, together with a dissipative coupling term associated with the collisions.
The generalized Gross-Pitaevskii equation derived in e.g. [\ref{[KD1]}], [\ref{[KD2]}], [\ref{[PBMR]}] and [\ref{[GNZ]}], is of the type
\begin{eqnarray}
i\hbar \partial _t\psi= &-\frac{\hbar^2}{2m}\Delta_x\psi+\Big( g|\psi|^2+U_{ext}+2g\int_{\R ^3} fdp+i\frac{g^2}{2{\hbar }}\int_{\R ^3\times \R ^3\times \R ^3} {A}\delta(p_1-p_2-p_3)\delta(E_1-E_2-E_3)\nonumber\\
&((1+f_1)f_2f_3-f_1(1+f_2)(1+f_3))dp_1dp_2dp_3\Big) \psi.
\end{eqnarray}
The two component problem (1.1), (1.4) is extensively discussed in the physics literature (see [\ref{[Eck]}],  [\ref{[HM]}],  [\ref{[IT]}], [\ref{[ITG]}], [\ref{[K]}], [\ref{[KD1]}], [\ref{[KD2]}], [\ref{[KK]}], [\ref{[PBMR]}], [\ref{[ST]}], [\ref{[ZNG]}]). It is proven in [\ref{[A]}] that these equations as given in [\ref{[PBMR]}], conserve the total energy. That is not so in some of the other settings, in particular not for (\ref{eq-f})-(\ref{eq-psi}) below.\\
\subsection{The model under study.}
We restrict to the 'high temperature range', and more particularly consider the temperature range close to $0.7T_{c}$.
As discussed in [\ref{[Eck]}], [\ref{[KD1]}], [\ref{[KD2]}],  [\ref{[ZNG]}] and more in details in [\ref{[ITG]}], then
$|p|\gg\sqrt{2mgn_c}$, the approximation $E(p)= \frac{ \lvert p\rvert ^2}{2m}+gn_c$ of the quasi-particle energy is  commonly used, $A= 1$, the operator $C_{22}$ is negligible, and the mass of the condensate exceeds that of the excitations,
 i.e. $n_c>\int P(p)dp$.
In equilibrium, the right hand side of (1.1) vanishes. Multiplying the collision term by $\log\frac{ f}{1+ f}$ and integrating in $p$, it follows that in equilibrium
\begin{eqnarray}
\frac{f_1}{1+f_1}=\frac{f_2}{1+f_2}\frac{f_3}{1+f_3}, \quad \quad \quad \quad {\rm when}\quad p_1=p_2+p_3, \quad \lvert p_1\rvert ^2= \lvert p_2\rvert ^2+\lvert p_3\rvert ^2+2mgn_c.\quad \quad \quad
\end{eqnarray}
Equation (1.5) implies that $\frac{f}{1+f}$ is a Maxwellian, hence the phase space density $f$ of the excitations is a Planckian, which is of the type
\begin{eqnarray*}
\frac{1}{e^{\alpha(\lvert p\rvert ^2+2mgn_c)+\beta\cdot p}-1}, \quad\alpha>0,\quad\beta\in\mathbb{R}^3,\quad p\in \R ^3.
\end{eqnarray*}
In the equilibrium Planckian distribution function, fix the condensate as identically equal to a constant $n_0>0$.  Set $\alpha =1$, take the $x$-component of $\beta $ as zero, $|\beta| =2\sqrt{2mgn_0}$ and write the Planckian as $\frac{1}{e^{\lvert p-p_0\rvert ^2}-1}$ with $p_0= -\frac{\beta}{2}$. Changing variables $p\rightarrow p-p_0$ gives
\begin{eqnarray*}
P(p):=\frac{1}{e^{\lvert p\rvert ^2}-1},\quad p\in \R ^3,
\end{eqnarray*}
as equilibrium Planckian distribution function. \\
The present paper studies the stability of the equilibrium $(P,\sqrt{n_0})$ of the system under small deviations, that respect the conservation laws. Although we are not deriving hydrodynamic limits, we take into account that the system is close to equilibrium and introduce a mean free path $\epsilon$, so that $C_{12}$ becomes $\frac{1}{\epsilon }C_{12}$. The factor $g$ is proportional to the scattering length $a$, which is smaller than the mean free path $\epsilon$. Take $\lambda $ of magnitude bounded by $(\frac{g}{\epsilon})^2(<1)$.
The functions $(f(t,x,p),\psi(t,x))$ are considered in the slab $\Omega = [ 0,2\pi ] $ in the $x$-direction with periodic boundary conditions, and taken as
\begin{eqnarray*}
(f,\psi )= (P(1+\lambda R), \sqrt{n}_0+\lambda \Phi ).
\end{eqnarray*}
In this paper the external potential $U_{ext}$ is assumed to be a constant that will be further discussed. We could alternatively have left out the external potential in (1.4) but replaced $\psi$ by $e^{itU_{ext}}\psi$ in the proofs. The atomic mass $m$ (resp. the Planck constant $\hbar $) will be taken as $\frac{1}{2}$ (resp. one) for simplicity.  Contrary to the classical Boltzmann operator in velocity space, $f\in L^1(\R ^3)$ does not imply $C_{12}(f)\in L^1(\R ^3)$. This paper is restricted to distribution functions, cylindrically symmetric in $p= (p_x,p_r)\in \R \times \R ^2$. That changes the linear moment conservation Dirac measure in the collision term to $\delta(p_{1x}-p_{2x}-p_{3x})$. Since the collective excitations play no role within the present temperature range, the domain of integration is here taken as the set of $p\in \R^3$ such that $\lvert p\rvert^2 >\Lambda^2 $ with $\Lambda \gg 2\sqrt{2gn_0}$. Denote by $\tilde{\chi}$ the characteristic function of the set
\begin{eqnarray*}
\{ (p,p_1,p_2,p_3)\in \R ^3\times \R ^3\times \R ^3\times\R^3;|p|^2, |p_1|^2,|p_2|^2, |p_3|^2>\Lambda^2\}.
\end{eqnarray*}
The restriction $|p|^2>\Lambda^2$ will be implicitly assumed below, and $\int dp$ will stand for $\int _{|p|^2>\Lambda^2} dp$. Set
\begin{eqnarray*}
\delta_3=\delta (p- p_1)-\delta (p- p_2)-\delta (p- p_3)\quad \text{and}\quad
\delta _0= \delta (p_{1x}= p_{2x}+p_{3x}, \hspace*{0.03in} \lvert p_1\rvert ^2= \lvert p_2\rvert ^2+\lvert p_3\rvert ^2+n_0).
\end{eqnarray*}
The system of equations to be satisfied by $(f,\psi )$ is
\hspace{1cm}\\
\begin{equation}\label{eq-f}
\partial _tf+p_x\partial _xf= g\sqrt{\lambda }n_c\int _{\R ^3\times \R ^3\times \R ^3}\tilde{\chi}\delta _0\delta _3(f_2f_3-f_1(1+f_2+f_3))dp_1dp_2dp_3,
\end{equation}
\begin{equation}\label{init-f}
f(0,x,p)=f_i(x,p),
\end{equation}
and
\begin{equation}\label{eq-psi}
\partial _t\psi -i\partial ^2_x\psi \\
= \Big( \frac{\sqrt{\lambda }}{2}\int_{\R ^3\times \R ^3\times \R ^3} \tilde{\chi}\delta _0(f_2f_3-f_1(1+f_2+f_3))dp_1dp_2dp_3-i(n_c+{\frac{U_{ext}}{g}}+2\int fdp)\Big) g\psi ,
\end{equation}
\begin{equation}\label{init-psi}
\psi (0,x)=\psi_i(x).
\end{equation}
Here, the function $n_c$ is defined by $n_c=n_c(t,x):= |\psi|^2(t,x)$. The approximate energy $\lvert p\rvert ^2+{g}n_c$ used in (1.5), at this range of temperature is replaced by $\lvert p\rvert ^2+gn_0$ as an approximation of order $\lambda $.\\
The total initial mass is
\begin{eqnarray*}
2\pi \mathcal{M}_0:= \int _\Omega |\psi_i(x)|^2dx+ \int  _{\Omega \times \R ^3}f_i(x,p)dxdp,
\end{eqnarray*}
which is formally conserved by the equations (1.6) and (1.8).\\
The initial data $f_i$ and $\psi _i$ are taken as
\begin{eqnarray*}
f_i:= P(1+\lambda R_i),\quad \psi_i:  = \sqrt{n_0}+\lambda \Phi _i,
\end{eqnarray*}
for some functions $R_i(x,p)$ and $\Phi _i(x)$ with
\begin{equation*}\label{cond2-Ri}
\int (|\psi_i|^2-n_0+\lambda \int_{\mathbb{R}^3}{P}R_idp)dx=0.
\end{equation*}
This is consistent with the asymptotic behavior proven in the paper, i.e. $(f-P,\lvert \psi \rvert ^2-n_0)$ tending to zero when time tends to infinity.
It implies that (up to the multiplicative constant $\frac{1}{2\pi }$) the initial (and conserved) total mass equals the mass of $(P,n_0)$, i.e.
\begin{equation}\label{cons-mass}
\mathcal{M}_0= \int P(p)dp+n_0.
\end{equation}
The separate masses of condensate and excitation may, however, not be conserved.
The constant $U_{ext}$ will be taken as ${g(n_0-2\mathcal{M}_0)}$. For a discussion of general modeling aspects, see also our paper [\ref{[AN1]}] and its references.\\
\hspace{1cm}\\
\subsection{The main mathematical result.}
The main results of the paper concern the well-posedness and long time behaviour of the problem (1.6-9). \\
For an initial perturbation of an equilibrium $(P,\sqrt{n}_0)$ of order $(\frac{g}{\epsilon})^2$ and conserving the total mass, the axial momentum and the kinetic energy of the excitations, the problem is well posed and the asymptotic limit when $t\rightarrow +\infty $ of the quasi-particle phase space density and the condensate mass are $P$ resp. $2\pi n_0$. The mass of the excitations together with the mass, the kinetic energy and the internal energy of the condensate converge exponentially to their equilibrium values when $t\rightarrow +\infty $. \\
Let $\parallel.\parallel_2$ denote the norm in $L^2([0,2\pi])$, and
set $\parallel\psi\parallel_{H^1}:=\parallel\psi\parallel_2+
\parallel\partial_x\psi\parallel_2$, let  $\parallel.\parallel_{2,2}$ denote the norm in $L^2_{\frac{P}{1+P}}([0,2\pi]\times\mathbb{R}^3)$, i.e.
\begin{eqnarray*}
\parallel h\parallel _{2,2}:= (\int h^2(x,p)\frac{P}{1+P}dpdx)^\frac{1}{2},
\end{eqnarray*}
and let $L^2_{\frac{1}{P(1+P)}}$ denotes the $L^2$-space of functions $h$ with norm $(\int \frac{h^2(x,p)}{P(1+P)}dpdx)^\frac{1}{2}$.\\
The solutions of (1.6-7) will be strong solutions, i.e. such that the collision operator $C_{12}(f,n_c)$ belongs to $C_b\big( \R ^+; L^2_{\nu ^{-\frac{1}{2}}\sqrt{\frac{P}{1+P}}}(\R ^3;H^1(0,2\pi ))\big) $, $\nu $ being the collision frequency defined in (\ref{df-nu}). The solutions of {(1.8-9)} are $H^1$-solutions in the following sense. A function $\psi\in \mathcal{C}_b(\mathbb{R}^+;H_{{\rm per}}^1(0,2\pi))$ is an $H^1$-solution to (1.8-9), if for all $\phi\in \mathcal{C}(\mathbb{R}^+;H_{{\rm per}}^1(0,2\pi))$ and all $t>0$,
\[ \begin{aligned}
&\int \psi(t,x)\bar{\phi}(t,x)dx-\int \psi_i(x)\bar{\phi}(0,x)dx+i\int_0^t\int \partial_x\psi(s,x)\partial_x\bar{\phi}(s,x) dxds\\
&=\int_0^t \int \Big( \frac{\sqrt{\lambda }}{2}\int_{\R ^3\times \R ^3\times \R ^3} \tilde{\chi}\delta _0(f_2f_3-f_1(1+f_2+f_3))dp_1dp_2dp_3-i(n_c+n_0-2\mathcal{M}_0+2\int fdp)\Big)g\psi\bar{\phi} dxds.
\end{aligned}\] \\
\begin{theorem}
\hspace*{0.1in}\\
There are $\lambda _1$, $c_\zeta $ and $\eta _1>0$, such that for $\lambda <\lambda _1$
and\\
$(R_i,\Phi _i)\in L_{(1+|p|)^3\frac{P}{1+P}}^2(\mathbb{R}^3;H^1_{{\rm per}}(0,2\pi))\times H_{{\rm per}}^1(0,2\pi)$ with
\begin{eqnarray}\label{cond1-Ri}
\int R_i(x,p)p_x{P}dxdp= \int R_i(x,p)(|p|^2+gn_0){P}dxdp= 0,
\end{eqnarray}
\begin{equation}\label{cond2-Ri}
\int (|\psi_i|^2-n_0+\lambda \int_{\mathbb{R}^3}{P}R_idp)dx=0,
\end{equation}
and
\begin{eqnarray}\label{cond3-Ri}
\parallel \Phi _i\parallel _{H^1}\leq \eta _1,\quad \parallel R_i\parallel _{2,2}+\parallel \partial _xR_i\parallel _{2,2}\leq \eta _1,
\end{eqnarray}
there is a unique solution
\begin{eqnarray*}
(f,\psi)=(P(1+\lambda R),\sqrt{n_0}+\lambda \Phi)\in  \mathcal{C}_b(\mathbb{R}^+;L^2_{\frac{1}{P(1+P)}}(\R^3;H_{{\rm per}}^1(0,2\pi)))\times \mathcal{C}_b(\mathbb{R}^+;H_{{\rm per}}^1(0,2\pi))
\end{eqnarray*}
to (1.6-9) with $f>0$. For all $t\in \mathbb{R}^+$, the solution satisfies,
\begin{eqnarray}\label{bounds-Phi-s}
f\in L^2_{\frac{(1+|p|)^3}{P(1+P)}}([0,t]\times\R^3;H_{{\rm per}}^1(0,2\pi))),\nonumber\\
\parallel R(t,\cdot ,\cdot )\parallel _{2,2}+\parallel \partial _xR(t,\cdot ,\cdot )\parallel _{2,2}\leq c_\zeta \eta _1e^{-\zeta t},\\
\int (\lvert \partial _x\psi\rvert ^2+\frac{g }{2}(\lvert \psi\rvert ^2-n_0)^2)(t,x)dx
\leq 2\lambda ,\nonumber
\end{eqnarray}
where $\zeta = c_\zeta \sqrt{\lambda }$. \\
Moreover, $n_c(t)=\int|\psi(t ,x)|^2dx$ converges exponentially of order $\zeta$ to $n_0$, when $t\rightarrow +\infty$,
\begin{equation}
\lim_{t\rightarrow +\infty}\int (\lvert \partial _x\psi\rvert ^2+\frac{g}{2}(\lvert \psi\rvert ^2-n_0)^2)(t,x)dx
\end{equation}
 exists, and the convergence to its limit is exponential of order $\zeta $.
\end{theorem}
\[\]
Whereas non-linear systems of the type (1.6-9) and its generalizations have been much studied in mathematical physics below $T_c$, there are so far only few papers with their focus mainly on the non-linear mathematical questions. Starting from a similar Gross-Pitaevskii and kinetic frame, two-fluid models are derived in [\ref{[A]}]. The space homogeneous initial value problem for this system is treated in [\ref{[AN1]}] for a large data setting. A Milne problem related to the present set-up is studied in [\ref{[AN2]}]. The paper [\ref{[EPV]}] considers a related setting,  and has its focus on linearized space homogeneous problems. Validation aspects in the space-homogeneous case are discussed in [\ref{[S]}].
There has also been a considerable interest recently (see ee.g. [\ref{[EV]}], [\ref{[L]}] and references therein) in the bosonic Nordheim-Uehling-Uhlenbeck equation as a model above and around $T_c$ for blow-ups and for condensation in space-homogeneous boson gases. \\
\\
A classical approach to study kinetic equations in a perturbative setting, is to use a spectral inequality (resp. Fourier techniques and the $\parallel \cdot \parallel _{T,2,2}$ norm) for controlling the non-hydrodynamic (resp. hydrodynamic) part
of a solution. An additional problem here is the coupling with the generalized Gross-Pitaevskii equation. The general approach, together with a Fourier based analysis of the generalized Gross-Pitaevskii equation, provide local in time solutions to the present coupled system.
Since the condensate and the normal gas are coupled by the collision interaction, the exponential decrease of
the deviation of the kinetic distribution function from the equilibrium Planckian $P$, helps to control the long-term evolution of the condensate. This is an important ingredient in the passage from local to global solutions,
 which leads to exponential decreases of the deviation of the condensate mass from its equilibrium state $n_0$, and of the energy (1.15) from its limit value.\\
Within this frame the kinetic equation (1.6) differs from earlier classical ones. The collision operator in space-homogeneous bosonic Nordheim-Uehling-Uhlenbeck papers has so far been taken isotropic, but is here, due to the space-dependent slab-context, cylindric. Mass density does not belong to the kernel of the present linearized collision operator. The scaling at infinity in its collision frequency is stronger than in the classical case.\\
\\
The one-dimensional spatial frame induces simplifications of the functional analysis, mainly in the control of the condensate. The $\T ^d$ spatial frame, for $d\geq 2$, is an open problem. \\
The conservation properties of the model (1.6-9), as well as some properties of the collision operator $\frac{C_{12}}{n_c}$ and its linearized operator $L$ around the Planckian $P$, are discussed in Section 2, including a spectral estimate for $L$. This is used in Section 3, which is devoted to a priori estimates for some linear equations related to (1.6) and (1.8). They are then employed in the proof of the main theorem in Section 4. The proof starts with a contractive iteration scheme to obtain local solutions. A key point in the global in time analysis is the exponential convergence to equilibrium for $f$ when $t\rightarrow +\infty $. The analysis of $\psi$ differs from the classical Gross-Pitaevskii case. It uses the exponential convergence to equilibrium of $f$ to control the behaviour of the kinetic energy $\int |\partial_x\psi|^2dx$ and the internal energy $\frac{g}{2}\int |\psi|^4dx$ of $\psi$.\\
%
%
%
%
%
%
\setcounter{equation}{0}
\setcounter{theorem}{0}
\section{Some properties of the model and the collision operator.}
The model induces total mass conservation as well as axial momentum and kinetic energy conservations for the excitations, as stated in the following lemma.
\begin{lemma}\label{conser}
It holds that
\begin{equation}\label{conser-mass}
\frac{d}{dt}\Big( \int _{\Omega \times \R ^3}f(t,x,p)dxdp+\int _\Omega \lvert \psi (t,x)\rvert ^2dx\Big) = 0,
\end{equation}
\begin{equation}\label{conser-momentum}
\frac{d}{dt}\int _{\Omega \times \R ^3}p_xf(t,x,p)dxdp= 0,
\end{equation}
\begin{equation}\label{conser-energy}
\frac{d}{dt}\int _{\Omega \times \R ^3}(\lvert p\rvert ^2+gn_0)f(t,x,p)dxdp= 0.
\end{equation}
\end{lemma}
\underline{Proof of Lemma \ref{conser}.}
\hspace*{0.1in}\\
Integrate (\ref{eq-f}) with respect to space and momentum. Add it to (\ref{eq-psi}) multiplied by $\bar{\psi }$ ( resp. the conjugate of (\ref{eq-psi})) multiplied by $\psi $) integrated with respect to space. One obtains (\ref{conser-mass}). Multiplying (\ref{eq-f}) by $p_x$ (resp. $(\lvert p\rvert ^2+gn_0$) and integrating it w.r.t. space and momentum leads to (\ref{conser-momentum}) (resp. (\ref{conser-energy})). \cqfd \\
\hspace*{1.in}\\
\hspace*{1.in}\\
Since the solutions will remain close to an equilibrium $(P,\sqrt{n_0})$, the linearized operator of $C_{12}$ around $P$ is of interest. For $\gamma := \sqrt{\lambda }$, consider the decomposition
\begin{eqnarray*}
f= P( 1+\gamma \tilde{R}) ,\quad \psi = \sqrt{n_0}+\gamma \tilde{\Phi }.
\end{eqnarray*}
It holds
\begin{equation*}
|\psi|^2= n_c=n_0+\gamma \sqrt{n_0}(\tilde{\Phi } +\bar{\tilde{\Phi }}) +\gamma ^2|\tilde{\Phi }|^2,
\end{equation*}
and the collision term can be written
\[ \begin{aligned}
\int \tilde{\chi}\delta _0\delta _3&(f_2f_3-f_1(1+f_2+f_3))dp_1dp_2dp_3= \gamma \Big( PL\tilde{R}+\gamma Q(\tilde{R},\tilde{R})\Big) ,
\end{aligned}\]
where
\begin{eqnarray*}
L\tilde{R}:= \frac{1}{P}\int \tilde{\chi}\delta(p_{1x}=p_{2x}+p_{3x})\delta(\lvert p_1\rvert ^2= \lvert p_2\rvert ^2+\lvert p_3\rvert ^2+{g}n_0)(\delta(p-p_1)-\delta(p-p_2)-\delta(p-p_3))\\
\Big[  -(1+P_2+P_3)P_1\tilde{R}_1+(P_3-P_1)P_2\tilde{R}_2+(P_2-P_1)P_3\tilde{R}_3\Big]dp_1dp_2dp_3,
\end{eqnarray*}
and
\begin{equation}\label{df-Q}
2Q(\tilde{R},\tilde{S}):= \int \tilde{\chi}\delta _0\delta _3\Big( P_2P_3(\tilde{R}_2\tilde{S}_3+\tilde{R}_3\tilde{S}_2)-P_1\tilde{R}_1(P_2\tilde{S}_2+P_3\tilde{S}_3)-P_1\tilde{S}_1(P_2\tilde{R}_2+P_3\tilde{R}_3)\Big) dp_1dp_2dp_3.
\end{equation}
We recall some properties about $L$ proved in [\ref{[AN2]}].
\begin{lemma}
$L$ is a self-adjoint operator in $L^2_{\frac{P}{1+P}}$.
Within the space of rotationally invariant distribution functions, its kernel is the subspace spanned by $(|p|^2+gn_0)(1+P)$ and $p_x(1+P)$.
\end{lemma}
\hspace*{1.in}\\
The operator $L$ splits into $K-\nu$, where
\begin{eqnarray}\label{df-nu}
\nu (p):= \int \tilde{\chi}\delta_0(1+P_2+P_3)dp_2dp_3+ 2\int \tilde{\chi}\delta_0(P_3-P_1)dp_1dp_3
\end{eqnarray}
and
\begin{eqnarray}\label{df-K}
Kh(p):= \frac{2}{P(p)}\Big( \int \tilde{\chi}\delta_0(P_3-P)P_2h_2dp_2dp_3
+ \int \tilde{\chi}\delta_0(1+P+P_3)P_1h_1dp_1dp_3\nonumber\\
+\int \tilde{\chi}\delta_0(P_1-P)P_3h_3dp_1dp_3\Big).
\end{eqnarray}
%
%
\begin{lemma}
The collision frequency $\nu $ satisfies
\begin{eqnarray}\label{bounds-nu}
\nu _0(1+|p|)^3\leq \nu (p)\leq \nu _1(1+|p|)^3, \quad p= (p_x,p_r)\in \R \times \R^+,
\end{eqnarray}
for some positive constants $\nu _0$ and $\nu_1$. The operator $K$ is compact from $L^2_{\nu \frac{P}{1+P}}$ into $L^2_{\nu ^{-1}\frac{P}{1+P}}$.
\end{lemma}
\hspace*{1.in}\\
Denote by $(\cdot , \cdot )$ the scalar product in $L^2_{\frac{P}{1+P}}$, and by $\tilde{P}$ the orthonormal projection on the kernel of $L$. Set $h_{\parallel}:=\tilde{P}h$ and $h_{\perp}:=(I-\tilde{P})h$.\\
%
%
\begin{lemma}
$L$ satisfies the spectral inequality,
\begin{equation}\label{spectral-in}
- (Lh, h)\geq c_0(\nu h_{\perp},h_{\perp}),\quad h\in L^2_{(1+|p|)^3\frac{P}{1+P}},
\end{equation}
with $c_0>0$.
\end{lemma}
We will also need an estimate for the quadratic collision operator $Q$.
\begin{lemma}
For cylindrically symmetric functions $(
\rm{g},h)\in L^2_{\nu\frac{P}{1+P}}\times L^2_{\frac{P}{1+P}}$ \\
(resp. $(\rm{g},h)\in L^2_{\frac{P}{1+P}}\times L^2_{\nu\frac{P}{1+P}}$), it holds
\begin{eqnarray*}
(\int \nu^{-1}\frac{P}{1+P}(\frac{Q(\rm{g},h)}{P})^2dp)^{\frac{1}{2}}\leq c \Big( \int \nu \frac{P}{1+P}\rm{g}^2(p)dp\int \frac{P}{1+P}h^2(p)dp\Big) ^{\frac{1}{2}},
\end{eqnarray*}
(resp.
\begin{eqnarray*}
(\int \nu^{-1}\frac{P}{1+P}(\frac{Q(\rm{g},h)}{P})^2dp)^{\frac{1}{2}}\leq c \Big( \int \nu \frac{P}{1+P}h^2(p)dp\int \frac{P}{1+P}\rm{g}^2(p)dp\Big) ^{\frac{1}{2}}).
\end{eqnarray*}
\end{lemma}
\underline{Proof.}
Considering cylindrically symmetric functions, we will use $\rm{g}= \rm{g}(p_x,p_r^2)$, $h= h(p_x,p_r^2)$. The theorem is a consequence of  the  following estimates for each of the terms of $\frac{Q(h,h)}{P}$. They are of the type
\[ \begin{aligned}
&Q_1(\rm{g},h)(p):= 2\int k_{1}(p,p_2)\rm{g}_2dp_2\hspace*{0.04in}\text{where}\hspace*{0.04in}
k_{1}(p,p_2):= P_2\int \delta (p_x=p_{2x}+p_{3x},|p|^2= |p_2|^2+|p_3|^2+\it{g}n_0)\frac{P_3h_3}{P}dp_3,\\
&\text{or}\\
&Q_2(\rm{g},h)(p):= 2\int k_{2}(p,p_2)\rm{g}_2dp_2\hspace*{0.04in}\text{where}\hspace*{0.04in}
k_{2}(p,p_2):= P_2h\int \delta (p_{1x}=p_{2x}+p_{x},|p_1|^2= |p_2|^2+|p|^2+\it{g}n_0)dp_1,\\
&\text{or}\\
&Q_3(\rm{g},h)(p):= 2\int k_{3}(p,p_2)\rm{g}_2dp_2\hspace*{0.04in}\text{where}\hspace*{0.04in}
k_{3}(p,p_2):= P_2h\int \delta (p_x=p_{2x}+p_{3x},|p|^2= |p_2|^2+|p_3|^2+\it{g}n_0)dp_3,\\
&\text{or}\\
&Q_4(\rm{g},h)(p):= 2\int k_{4}(p,p_1)\rm{g}_1dp_1\hspace*{0.04in}\text{where}\hspace*{0.04in}
k_{4}(p,p_1):= P_1h\int \delta (p_{1x}=p_{x}+p_{3x},|p_1|^2= |p|^2+|p_3|^2+\it{g}n_0)dp_3,\\
&\text{or}\\
&Q_5(\rm{g},h)(p):= 2\int k_{5}(p,p_3)h_3dp_3\hspace*{0.04in}\text{where}\hspace*{0.04in}
k_{5}(p,p_3):= \frac{P_3}{P}\int \delta (p_{1x}=p_x+p_{3x},|p_1|^2=|p|^2+|p_3|^2+\it{g}n_0)P_1\rm{g}_1dp_1.
\end{aligned}\]
Let $(\rm{g},h)\in L^2_{\nu\frac{P}{1+P}}\times L^2_{\frac{P}{1+P}}$. Consider first the term $(\int \nu^{-1}\frac{P}{1+P}(\frac{Q_1(\rm{g},h)}{P})^2dp)^{\frac{1}{2}}$. $P$ is uniformly bounded by $M$ from above and below in the domain of integration, so in the estimates below it is enough to use $M$ instead of $P$. It holds
\[ \begin{aligned}
(\int \nu^{-1}M&(\int k_1(p,p_2)\rm{g}_2dp_2)^2dp)^{\frac{1}{2}}\\
&\leq
\int (\int \nu^{-1}Mk_1^2(p,p_2)dp)^{\frac{1}{2}}\rm{g}_2dp_2\\
&\leq c\int M_2\rm{g}_2\Big( \int \nu ^{-1}M^{-1}(Mh)^2(p_x-p_{2x},|p|^2-|p_2|^2-\it{g}n_0-|p_x-p_{2x}|^2)dp\Big) ^{\frac{1}{2}}dp_2\\
&\leq c\int M_2\rm{g}_2\Big( \int \nu ^{-1}(\sqrt{|p_2|^2+|p_3|^2+\it{g}n_0})M_2^{-1}M_3^{-1}(M_3h_3)^2)dp_3\Big) ^{\frac{1}{2}}dp_2\hspace{2cm}\\
&\leq c(\int \frac{1}{(1+\lvert p_2\rvert )^\frac{3}{2}}M_2^{\frac{1}{2}}\rm{g}_2dp_2)(\int M_3h_3^2dp_3)^{\frac{1}{2}}\\
&\leq c(\int \nu _2M_2\rm{g}_2^2dp_2)^{\frac{1}{2}}(\int M_3h_3^2dp_3)^{\frac{1}{2}},
\end{aligned}\]
by the Cauchy-Schwartz inequality. For the $(\int \nu^{-1}\frac{P}{1+P}(\frac{Q_2(\rm{g},h)}{P})^2dp)^{\frac{1}{2}}$ term,
\[\begin{aligned}
(\int \nu^{-1}M(\int k_2(p,p_2)\rm{g}_2dp_2)^2dp)^{\frac{1}{2}}&\leq
\int (\int \nu^{-1}Mk_2^2(p,p_2)dp)^{\frac{1}{2}}\rm{g}_2dp_2\\
&\leq c(\int M_2\rm{g}_2dp_2)(\int \nu^{-1}Mh^2dp)^{\frac{1}{2}}\\
&\leq c(\int \nu _2M_2\rm{g}_2^2dp_2)^{\frac{1}{2}}(\int Mh^2dp)^{\frac{1}{2}}.
\end{aligned}\]
The $\Big( (\int \nu^{-1}\frac{P}{1+P}(\frac{Q_i(\rm{g},h)}{P})^2dp)^{\frac{1}{2}}\Big) _{3\leq i\leq 4}$ terms can be handled similarly. Finally,
\[ \begin{aligned}
(\int \nu^{-1}M(\int \tilde{k}_5(p,p_3)h_3dp_3)^2dp)^{\frac{1}{2}}&\leq \int (\int \nu^{-1}M\tilde{k}_5^2(p,p_3)dp)^{\frac{1}{2}}h_3dp_3\\
&\leq c\int M_3^{\frac{3}{2}}h_3dp_3\Big( \int M_1\rm{g}_1^2dp_1\Big) ^{\frac{1}{2}} \\
&\leq c(\int M_3h_3^2dp_3)^{\frac{1}{2}}(\int M_1\rm{g}_1^2dp_1)^{\frac{1}{2}}.
\end{aligned}\]
This completes the proof of the lemma.
\cqfd
\begin{lemma}
\hspace*{0.1in}\\
There is a constant $c>0$ such that for any cylindrically symmetric function $f\in L^2(\R ^3)$,
\begin{eqnarray*}
\mid \int P(Lf)(p)dp\mid \leq c\Big( \int \frac{P}{1+P}f_\perp ^2(p)dp\Big) ^{\frac{1}{2}}.
\end{eqnarray*}
\end{lemma}
\underline{Proof.} Using the Cauchy-Schwartz inequality,
\[ \begin{aligned}
\mid \int P(Lf)(p)dp\mid &=  \mid \int \big( \sqrt{P}\nu \big) \big( \frac{\sqrt{P}}{\nu }Lf\big) dp\mid \\
&\leq c\Big( \int Pf_\perp ^2(p)dp\Big) ^{\frac{1}{2}}\\
&\leq c\Big( \int \frac{P}{1+P}f_\perp ^2(p)dp\Big) ^{\frac{1}{2}}.
\end{aligned}\]
\hspace*{0.1in}\\
\hspace*{0.1in}\\
%
%
%
%
%
%
%
%
\section{Rest term estimates.}
\setcounter{equation}{0}
\setcounter{theorem}{0}
Consider the decomposition
\begin{eqnarray*}
f= P( 1+\gamma \tilde{R}) ,\quad \psi = \sqrt{n_0}+\gamma \tilde{\Phi }.
\end{eqnarray*}
The equations (\ref{eq-psi})-(\ref{init-psi}) written for $\tilde{\Phi }$ with periodic boundary conditions when $\tilde{R}$ is given, are
\begin{equation}\label{eq-Phi}
\partial_t\tilde{\Phi }-i\partial^2_x\tilde{\Phi } = S_1\tilde{\Phi }+{S_2}\bar{\tilde{\Phi }}+U,\quad \quad \tilde{\Phi} (0,\cdot )= \tilde{\Phi} _i.
\end{equation}
Here $S_1$ and ${S_2}$ are the coefficients of the linear terms in $\tilde{\Phi }$ resp. $\bar{\tilde{\Phi }}$, and $U$ contains the inhomogeneous terms and the non-linear terms in $\tilde{\Phi }$, $\bar{\tilde{\Phi }}$. In the following lemmas the dependence of $U$ on $\tilde{\Phi }$ is not taken into account.
\begin{lemma}\label{1st-control}
\hspace*{0.1in}\\
Let $\tilde{\Phi }_i$ (resp. $S_1$, $S_2$, $U$) be a given function in $H^1_{per}(0,2\pi )$ (resp. $L^{\infty} (\R ^+;H^1_{per}(0,2\pi ))$. \\
There is a unique solution $\tilde{\Phi }$ to (3.1) in $\mathcal{C}(\R ^+; H^1_{per}(0,2\pi ))$. Moreover,
\begin{equation}\label{control-Phi}
\parallel \tilde{\Phi }(t,.)\parallel^2_{H^1}\leq (2\parallel \tilde{\Phi }_i\parallel^2_{H^1}+6t\int_0^t\parallel U(s,.)\parallel^2_{H^1}ds)e^{6t\int_0^t(\parallel {S_1}(r,.)\parallel^2_{H^1}+\parallel {S_2}(r,.)\parallel^2_{H^1})dr},\quad t>0.
\end{equation}
\end{lemma}
\hspace{1cm}\\
\underline{Proof of Lemma 3.1}\\
Consider first the equations
\begin{equation}\label{eq-Phi-W}
\partial_t\tilde{\Phi }-i\partial^2_x\tilde{\Phi } = W,\quad \quad \tilde{\Phi } (0,\cdot )= \tilde{\Phi } _i,
\end{equation}
for a given $W\in L^{\infty} (\R ^+;H^1_{per}(0,2\pi ))$. Writing $W$ and $\tilde{\Phi }$ in Fourier
series, gives
\begin{eqnarray*}
\hat{\tilde{\Phi }}^\prime _n(t)+in^2\hat{\tilde{\Phi }}_n=\hat{W}_n,
\end{eqnarray*}
and so
\begin{equation}\label{Fourier-Phi}
\hat{\tilde{\Phi }}_n(t)=\hat{\tilde{\Phi }}_n(0)e^{-in^2t}+\int_0^t \hat{W}_n(s)e^{in^2(s-t)}ds.
\end{equation}
Hence,
\[ \begin{aligned}
&|\hat{\tilde{\Phi }}_n(t)|\leq |\hat{\tilde{\Phi }}_n(0)|+\int_0^t |\hat{W}_n(s)|ds,\\
&\sum _{n\in \N } (1+n^2)|\hat{\tilde{\Phi }}_n(t)|^2\leq 2\sum _{n\in \N } [(1+n^2) |\hat{\tilde{\Phi }}_n(0)|^2+t\int_0^t (1+n^2) |\hat{W}_n(s)|^2ds].\\
\end{aligned}\]
And so the function $\tilde{\Phi }$ defined by (\ref{Fourier-Phi}) belongs to $L^{\infty}_{loc} (\R ^+;H^1_{per}(0,2\pi ))$. Moreover,
\[ \begin{aligned}
&\parallel \tilde{\Phi }(t,.)\parallel^2_{H^1}\leq 2\Big( \parallel \tilde{\Phi }_i\parallel^2_{H^1}+t\int_0^t \parallel W(s,.)\parallel^2_{H^1}ds\Big) .
\end{aligned}\]
We conclude that given $\tilde{\Phi }_i\in H^1_{per}(0,2\pi )$ and $W\in L^{\infty} (\R ^+;H^1_{per}(0,2\pi ))$, there exists a unique solution
$\tilde{\Phi }\in L^{\infty}_{loc} (\R ^+;H^1_{per}(0,2\pi ))$ to (\ref{eq-Phi-W}). It also follows from (\ref{Fourier-Phi}) that the solution is a continuous function of $t\in \mathbb{R}^+$ into $H^1_{per}(0,2\pi )$.
For $W= W(\tilde{\Phi }):= S_1\tilde{\Phi }+{S_2}\bar{\tilde{\Phi }}+U $ it holds,
\[ \begin{aligned}
\parallel W(s,.)\parallel^2_{H^1}&\leq 3\parallel S_1\tilde{\Phi }(s,.)\parallel^2_{H^1} +3\parallel{S_2}\bar{\tilde{\Phi }}(s,.)\parallel^2_{H^1}+3\parallel U(s,.)\parallel^2_{H^1}\\
&\leq 3\parallel S_1(s,.)\parallel^2_{H^1}\parallel \tilde{\Phi }(s,.)\parallel^2_{H^1} +3\parallel{S_2}(s,.)\parallel^2_{H^1}\parallel\bar{\tilde{\Phi }}(s,.)\parallel^2_{H^1}+3\parallel U(s,.)\parallel^2_{H^1}.
\end{aligned}\]
With $\tilde{\Phi }_0=0$, an iterative sequence of solutions $\tilde{\Phi }_j$ of (\ref{eq-Phi-W}) for $j\geq 1$ with the right hand side $W(\tilde{\Phi }_{j-1})$, gives
\begin{equation}\label{Phi-without-Gronwall}
\parallel \tilde{\Phi }_j(t,.)\parallel^2_{H^1}\leq 2\parallel \tilde{\Phi }_i\parallel^2_{H^1}\\
+6t\int_0^t \Big( (\parallel S_1(s,.)\parallel^2_{H^1}+\parallel{S_2}(s,.)\parallel^2_{H^1})\parallel \tilde{\Phi }_{j-1} (s,.)\parallel^2_{H^1}+\parallel U(s,.)\parallel^2_{H^1}\Big) ds,
\end{equation}
and with $\delta \tilde{\Phi }_j= \tilde{\Phi }_j-\tilde{\Phi }_{j-1}$,
\begin{equation*}
\parallel\delta \tilde{\Phi }_j(t,.)\parallel^2_{H^1}\leq
6t\int_0^t \Big( (\parallel S_1(s,.)\parallel^2_{H^1}+\parallel{S_2}(s,.)\parallel^2_{H^1})\parallel \delta\tilde{\Phi }_{j-1} (s,.)\parallel^2_{H^1}.
\end{equation*}
It follows that the sequence converges on some interval $t\in [0,T]$, and that (\ref{Fourier-Phi}) and (\ref{Phi-without-Gronwall}) hold for the limit $\tilde{\Phi }$, a unique solution of (\ref{eq-Phi}). By an iteration of the argument the existence and the continuity of $\tilde{\Phi }$ hold for $t>0$.
Using Gronwall on (\ref{Phi-without-Gronwall}) for $\tilde{\Phi }$ gives (3.2).
\cqfd
\[\]
\hspace{1cm}\\
The rest of this section prepares for the control of the excitation distribution function $f$ around the equilibrium $P$. With
\begin{eqnarray*}
f= P( 1+\gamma \tilde{R}) ,\quad f_i= P( 1+\gamma \tilde{R}_i),
\end{eqnarray*}
the equations (\ref{eq-f})-(\ref{init-f}) written for $\tilde{R}$, are
\begin{equation}\label{R}
\partial_t\tilde{R}+p_x\partial_x\tilde{R}= g\gamma \big( n_0L\tilde{R}+\gamma(L_1\tilde{R}+Q_1(\tilde{R},\tilde{R}))\big) ,\quad \tilde{R}(0,\cdot ,\cdot )= \tilde{R}_i,
\end{equation}
where $L_1$ (resp. $Q_1$) is a linear (resp. quadratic) operator. \\
The following norms are used. For $1\leq q\leq \infty$,
\[ \begin{aligned}
&\parallel f \parallel_{2,q}=\Big(\int_{\mathbb{R}^3}\frac{P}{1+P}
(\int_{[0,2\pi]}|f(x,p)|^qdx)^{\frac{2}{q}}dp\Big)^{\frac{1}{2}},\\
&\parallel f \parallel_{2,H^1}=\Big(\int_{ [ 0,2\pi ] \times \mathbb{R}^3}\frac{P}{1+P}\big( |f(x,p)|^2+
|\partial_xf(x,p)|^2\big) dxdp\Big)^{\frac{1}{2}},\\
&\parallel f \parallel_{T,2,2}=\Big(\int_{ [0,T] \times [ 0,2\pi ] \times \mathbb{R}^3}\frac{P}{1+P}
|f(t,x,p)|^2dtdxdp\Big)^{\frac{1}{2}}.\\
\end{aligned}\]
%
%
%
%
%
%
To study (\ref{R}), some a priori estimates will be needed for the linear problem
\begin{equation}\label{h}
{\partial_th+p_x\partial_xh= g\gamma \big( n_0 Lh+\gamma G\big) },\hspace{.5cm}h(0,\cdot ,\cdot )= h_0,
\end{equation}
periodic in $x$ with period $2\pi$. Assume $G\in L^{\infty}([0,T];L_{(1+|p|)^{-3}\frac{P}{1+P}}^2(\mathbb{R}^3;H^1_{{\rm per}}(0,2\pi)))$ for $T>0$. The function $\partial_x h$ is at least formally a solution to
\begin{equation}\label{eq-derivative-h}
\partial_t(\partial _xh)+p_x\partial_x(\partial _xh)= g\gamma\big( n_0 L(\partial _xh)+\gamma \partial _xG\big) ,\hspace{.5cm}\partial _xh(0,\cdot ,\cdot )= \partial _xh_0,
\end{equation}
periodic in $x$ with period $2\pi$.
For existence of solutions to problems of the type (\ref{h}), see [\ref{[M]}] or alternatively, consider the Fourier transform in $x$ of (\ref{h}) and argue as in the proof of Lemma 3.1. The solutions are unique and continuous as functions of $t$ into $L_{\frac{P}{1+P}}^2(\mathbb{R}^3;H^1_{{\rm per}}(0,2\pi))$.
Multiply the equation by $h\frac{P}{1+P}$, integrate on $[0,{T}]\times [0,2\pi]\times \mathbb{R}^3$,  and use (\ref{spectral-in}) to get
\begin{lemma}\label{control-h_perp}
For any $\eta >0$,
\begin{eqnarray*}
\parallel h(T,.)\parallel^2_{2,2}+\gamma \parallel \nu^{\frac{1}{2}} h_{\perp}\parallel^2_{T,2,2}\leq c(\parallel h_0\parallel^2_{2,2}+\gamma ^3
\parallel\nu^{-\frac{1}{2}}G_{\perp}\parallel^2_{T,2,2}+\gamma ^2\eta \parallel G_{\parallel}\parallel^2_{T,2,2}+\frac{\gamma ^2}{\eta }\parallel h_{\parallel}\parallel^2_{T,2,2}).
\end{eqnarray*}
\end{lemma}
\hspace{1cm}\\
\begin{lemma}\label{control-h_parallel}
Assume $\gamma \leq 1$,
\begin{equation}\label{hypG}
\int _0^{2\pi }\int P(\mid p\mid ^2+gn_0)G(t,x,p)dpdx= \int _0^{2\pi }\int Pp_xG(t,x,p)dpdx= 0,\quad t\in (0,T),
\end{equation}
and
\begin{equation}\label{hyp-h0}
\int _0^{2\pi }\int P(\mid p\mid ^2+gn_0)h_0(x,p)dpdx= \int _0^{2\pi }\int Pp_xh_0(x,p)dpdx= 0.
\end{equation}
Then
\begin{eqnarray}
\parallel {h}_{\parallel}\parallel_{T,2,2}^2\leq c\Big( \gamma ^{-1}\parallel h_0\parallel^2_{2,2}+
\gamma ^2\parallel \nu ^{-\frac{1}{2}}G_{\perp}\parallel^2_{T,2,2}+\gamma \parallel G_\parallel \parallel _{T,2,2}^2 \Big).\hspace{1cm}
\end{eqnarray}
\end{lemma}
\underline{Proof.}\\
Let $\{ \chi_x=c_xp_x(1+P)$, $\chi_2=c_2(|p|^2+1)(1+P)\}$ be an orthonormal basis for the kernel of $L$. Let $\chi _1= c_1(1+P)$ with $c_1= (\int P(1+P)dp)^{-\frac{1}{2}}$. Consider the Fourier series in $x$ of (3.7),
\begin{equation}\label{fourier-hk}
\partial _th_k+ikp_xh_k= c_k+g\gamma ^2G_k,
\end{equation}
where $h_k$ (resp. $c_k$, resp. $G_k$) is the $k$-th Fourier coefficient of $h$ (resp. $g\gamma n_0Lh$, resp. $G$). Set $h_{k2}= (h_k, \chi _2)$, $h_{kx}= (h_k, \chi _x)$ and $h_{k1}= (h_k,\chi _1)$. Since $\chi _1= \alpha \chi_2+\chi_{1\perp}$, where $\alpha = (\chi _1, \chi _2)$ and $\chi_{1\perp}$ is orthonormal to $\chi_2$ and $\chi_x$, it holds that $h_{k1}= \alpha h_{k2}+h_{k\chi_{1 \perp}}$. \\
Multiply (\ref{fourier-hk}) by $\frac{P}{1+P}\chi_2$ (resp. $\frac{P}{1+P}\chi_x$, resp. $\frac{P}{1+P}\chi_1$) and integrate in $p$,
 \begin{equation}\label{eq-h_k2}
 \partial_t h_{k2}+ik\kappa(h_{kx}+\kappa^{-1}h_{kp_x\chi_2})= g\gamma ^2G_{k2},
 \end{equation}
 \begin{equation}\label{eq-h_kx}
 \partial_t h_{kx}+ik\kappa(h_{k2}+\kappa^{-1}h_{kp_x\chi_x})= g\gamma ^2G_{kx},
 \end{equation}
  \begin{equation}\label{eq-h_k1}
 \partial_t h_{k1}+ik\frac{c_1}{c_x}h_{kx}= d_{k1}+g\gamma ^2G_{k1}.
 \end{equation}
 Here $h_{p_x\chi _x}$ and $h_{p_x\chi _2}$, denote non-hydrodynamic moments of h, $\kappa := \int \frac{P}{1+P}p_x\chi _2\chi _xdp$ and \\
 $d_1= \gamma (Lh, \chi _1)$.
Take the real ($\mathcal{R}$) and imaginary ($\mathcal{I}$ with $i$ included) parts of (\ref{eq-h_k2})-(\ref{eq-h_kx})-(\ref{eq-h_k1}), so that
 \begin{align*}
 &\partial_t \mathcal{R}h_{k2}+ik\kappa(\mathcal{I}h_{kx}+\kappa^{-1}
 \mathcal{I}h_{kp_x\chi_2})= g\gamma ^2\mathcal{R}G_{k2},\numberthis \label{Rhk2}\\
&\partial_t \mathcal{I}h_{k2}+ik\kappa(\mathcal{R}h_{kx}+\kappa^{-1}
 \mathcal{R}h_{kp_x\chi_2})= g\gamma ^2\mathcal{I}G_{k2},\numberthis \label{Ihk2}\\
&\partial_t \mathcal{R}h_{kx}+ik\kappa(\mathcal{I}h_{k2}+\kappa^{-1}\mathcal{I}h_{kp_x\chi_x})= g\gamma ^2\mathcal{R}G_{kx},\numberthis \label{Rhkx}\\
&\partial_t \mathcal{I}h_{kx}+ik\kappa(\mathcal{R}h_{k2}+\kappa^{-1}\mathcal{R}h_{kp_x\chi_x})= g\gamma ^2\mathcal{I}G_{kx},\numberthis \label{Ihkx}\\
&\partial_t \mathcal{R}h_{k1}+ik\frac{c_1}{c_x}\mathcal{I}h_{kx}
 = \mathcal{R}d_{k1}+\gamma ^2\mathcal{R}G_{k1},\numberthis \label{Rhk1}\\
&\partial_t \mathcal{I}h_{k1}+ik\frac{c_1}{c_x}\mathcal{R}h_{kx}
 =\mathcal{I}d_{k1}+ \gamma ^2\mathcal{I}G_{k1}.\numberthis \label{Ihk1}
 \end{align*}
Multiply (\ref{Rhk2}), (\ref{Ihk2}), (\ref{Rhkx}), (\ref{Ihkx}) by respectively $-i\mathcal{I}h_{kx}, i\mathcal{R}h_{kx}, i\mathcal{I}h_{k2}, -i\mathcal{R}h_{k2}$, and sum. This leads to
 \begin{align*}
\mid h_{kx}\mid ^2-\mid h_{k2}\mid ^2&= \frac{i}{k\kappa } \partial_t(\mathcal{R}h_{kx}\mathcal{I}h_{k2}-\mathcal{R}h_{k2}\mathcal{I}h_{kx})\\
&+\frac{1}{\kappa }\Big( \mathcal{I}h_{kx}\mathcal{I}h_{kp_x\chi _2}-\mathcal{R}h_{kx}\mathcal{R}h_{kp_x\chi _2}
 -\mathcal{I}h_{k2}\mathcal{I}h_{kp_x\chi _x}
 +\mathcal{R}h_{k2}\mathcal{R}h_{kp_x\chi _x}\Big) \\
 &+\frac{ig\gamma ^2}{k\kappa }\Big( \mathcal{I}h_{kx}\mathcal{R}G_{k2}-\mathcal{R}h_{kx}\mathcal{I}G_{k2}-\mathcal{I}h_{k2}\mathcal{R}G_{kx}+\mathcal{R}h_{k2}\mathcal{I}G_{kx}\Big) ,\quad k\neq 0.\numberthis \label{1st-difference}
 \end{align*}
Multiply (\ref{Rhk1}), (\ref{Ihk1}), (\ref{Rhkx}), (\ref{Ihkx}) by respectively $-i\mathcal{I}h_{kx}, i\mathcal{R}h_{kx}, i\mathcal{I}h_{k1}, -i\mathcal{R}h_{k1}$, and sum. This leads to
 \begin{align*}
\frac{c_1}{c_x}\mid h_{kx}\mid ^2-\alpha \kappa \mid h_{kx}\mid ^2&= \frac{i}{k}\partial_t(\mathcal{R}h_{kx}\mathcal{I}h_{k1}-\mathcal{R}h_{k1}\mathcal{I}h_{kx})\\
&+\frac{i}{k}\big( \mathcal{I}h_{kx}(\mathcal{R}d_{k1}+\gamma ^2\mathcal{R}G_{k1}) -\mathcal{R}h_{kx}( \mathcal{I}d_{k1}+\gamma ^2\mathcal{I}G_{k1})\big) \\
 &-\kappa \mathcal{I}h_{k\chi _1\perp }\mathcal{I}h_{k2}-\mathcal{I}h_{k1}\mathcal{I}h_{kp_x\chi _x}+\kappa \mathcal{R}h_{k\chi _1\perp }\mathcal{R}h_{k2}+\mathcal{R}h_{k1}\mathcal{R}h_{kp_x\chi _x}\\
 &+\frac{ig\gamma ^2}{k}\Big( \mathcal{R}h_{k1}\mathcal{I}G_{kx}-\mathcal{I}h_{k1}\mathcal{R}G_{kx}\Big) ,\quad k\neq 0.\numberthis \label{2nd-difference}
 \end{align*}
 Moreover, $\frac{c_1}{c_x}\neq \alpha \kappa $. Indeed, 
 \begin{align*}
 &\frac{c_1-c_x\alpha \kappa }{c_1c_x^2c_2^2}=\\
 &\int _{ |p|>\Lambda }P(1+P)p_x^2dp\int _{ |p|>\Lambda }P(1+P)(|p|^2+1)^2dp-\int _{ |p|>\Lambda }P(1+P)(|p|^2+1)dp\int _{ |p|>\Lambda }P(1+P)p_x^2(|p|^2+1)dp\\
 &= \int _{ |p|>\Lambda }P(1+P)p_x^2dp\int _{ |p|>\Lambda }P(1+P)(|p|^2+1)|p|^2dp-\int _{ |p|>\Lambda }P(1+P)(|p|^2+1)dp\int _{ |p|>\Lambda }P(1+P)p_x^2|p|^2dp\\
 &= 3(\int _{ |p|>\Lambda }P(1+P)p_x^2dp)^2-\int _{ |p|>\Lambda }P(1+P)dp(\int _{ |p|>\Lambda }P(1+P)p_x^4dp+2\int_{ |p|>\Lambda } P(1+P)p_x^2p_y^2dp)\\
 &= \frac{4\pi ^2}{3}\Big( \big( \Lambda ^3P(\Lambda )+3\int _{ \Lambda }^{+\infty }P(r)r^2dr\big) ^2-\big( \Lambda P(\Lambda )+\int _{\Lambda }^{+\infty }P(r)dr\big) \big( \Lambda ^5P(\Lambda )+5\int _{\Lambda }^{+\infty }P(r)r^4dr\big) \Big) \\
 &= \frac{4\pi ^2}{3}\Big( \big( \Lambda ^3e^{-\Lambda ^2}+3\int _{ \Lambda }^{+\infty }e^{-r^2}r^2dr\big) ^2-\big( \Lambda e^{-\Lambda ^2}+\int _{\Lambda }^{+\infty }e^{-r^2}dr\big) \big( \Lambda ^5e^{-\Lambda ^2}+5\int _{\Lambda }^{+\infty }e^{-r^2}r^4dr\big) +R_\Lambda \Big) ,
 \end{align*}
 where, with $M(r)= e^{-r^2}$, $R_\Lambda $ is defined by
 \begin{align*}
 R_\Lambda =&\big( \Lambda ^3(P-M)(\Lambda )+3\int _{ \Lambda }^{+\infty }(P-M)(r)r^2dr\big) \big( \Lambda ^3(P+M)(\Lambda )+3\int _{ \Lambda }^{+\infty }(P+M)(r)r^2dr\big)  \\
 &-\big( \Lambda (P-M)(\Lambda )+\int _{\Lambda }^{+\infty }(P-M)(r)dr\big) \big( \Lambda ^5M(\Lambda )+5\int _{\Lambda }^{+\infty }M(r)r^4dr\big) \\
 &-\big( \Lambda P(\Lambda )+\int _{\Lambda }^{+\infty }P(r)dr\big) \big( \Lambda ^5(P-M)(\Lambda )+5\int _{\Lambda }^{+\infty }(P-M)(r)r^4dr\big) .
 \end{align*}
 Consequently,
 \begin{align*}
 &\frac{c_1-c_x\alpha \kappa }{c_1c_x^2c_2^2}=\\
 &= \frac{4\pi ^2}{3}\Big( \big( (\Lambda ^3+\frac{3}{2}\Lambda )e^{-\Lambda ^2}+\frac{3}{2}\int _{ \Lambda }^{+\infty }e^{-r^2}dr\big) ^2\\
 &\hspace*{0.6in}-\big( \Lambda e^{-\Lambda ^2}+\int _{\Lambda }^{+\infty }e^{-r^2}dr\big) \big( (\Lambda ^5+\frac{5}{2}\Lambda ^3+\frac{15}{4}\Lambda )e^{-\Lambda ^2}+\frac{15}{4}\int _{\Lambda }^{+\infty }e^{-r^2}dr\big) +R_\Lambda \Big) \\
 &= \frac{4\pi ^2}{3}\Big( \frac{\Lambda ^2}{2}(\Lambda ^2-3)e^{-2\Lambda ^2}-\Lambda (\Lambda ^4-\frac{\Lambda ^2}{2}+3)e^{-\Lambda ^2}\int _{ \Lambda }^{+\infty }e^{-r^2}dr-\frac{3}{2}\big( \int _{ \Lambda }^{+\infty }e^{-r^2}dr\big) ^2+R_\Lambda \Big) .
 \end{align*}
But
\begin{align*}
\int _{ \Lambda }^{+\infty }\Lambda e^{-r^2}dr&= \frac{1}{2}\int _{\Lambda ^2}^{+\infty }\frac{\Lambda e^{-u}}{\sqrt{u}}du= \frac{e^{-\Lambda ^2}}{2}\int _0^{+\infty }\frac{\Lambda e^{-s}}{\sqrt{s+\Lambda ^2}}ds=\frac{e^{-\Lambda ^2}}{2}\int _0^{+\infty }(1+\frac{s}{\Lambda^2})^{-\frac{1}{2}}e^{-s}ds\\
 &\geq \frac{e^{-\Lambda ^2}}{2}\int _0^{+\infty }(1-\frac{s}{2\Lambda^2}+\frac{3}{8}\frac{s^2}{\Lambda^4}-\frac{5}{16}\frac{s^3}{\Lambda^6})e^{-s}ds\\
&= \frac{e^{-\Lambda ^2}}{2}\Big( 1-\frac{1}{2\Lambda^2}+\frac{3}{4\Lambda^4}-\frac{15}{8\Lambda^6}\Big) .
 \end{align*}
 And so,
 \begin{align*}
3e^{2\Lambda ^2}\frac{c_1-c_x\alpha \kappa }{4\pi ^2c_1c_x^2c_2^2}&\leq \frac{\Lambda^4}{2}-\frac{3}{2}\Lambda ^2
-(\Lambda ^4-\frac{\Lambda ^2}{2}+3) \Big( \frac{1}{2}-\frac{1}{4\Lambda ^2}+\frac{3}{8\Lambda ^4}-\frac{15}{16\Lambda ^6}\Big) +R_\Lambda e^{2\Lambda ^2}\\
&= -\Lambda ^2-2+\frac{15}{8\Lambda ^2}-\frac{51}{32\Lambda ^4}+\frac{45}{16\Lambda ^6}+R_\Lambda e^{2\Lambda ^2}.
\end{align*} 
The sum of the terms before $R_\Lambda e^{2\Lambda ^2}$ is neggative for $\Lambda \geq 2$. Moreover $R_\Lambda e^{2\Lambda ^2}$ is of order $e^{-\Lambda^2}$. Indeed,
 it follows from
 \begin{eqnarray*}
  \int _{ \Lambda }^{+\infty }e^{-r^2}dr\leq \frac{\sqrt{\pi }}{4}e^{-\Lambda ^2},
 \end{eqnarray*}
 and
 \begin{eqnarray*}
 0\leq (P-M)(r)\leq \frac{e^{-\Lambda ^2}}{1-e^{-\Lambda ^2}}e^{-r^2},\quad \quad \quad (P+M)(r)\leq \frac{2-e^{-\Lambda ^2}}{1-e^{-\Lambda ^2}}e^{-r^2},\quad r\geq \Lambda ,
 \end{eqnarray*}
 that
  \begin{align*}
 |R_\Lambda |\leq & \hspace*{0.1in}e^{-\Lambda ^2}\frac{2-e^{-\Lambda ^2}}{(1-e^{-\Lambda ^2})^2}\Big( \big( \Lambda ^3e^{-\Lambda ^2}+3\int _{ \Lambda }^{+\infty }e^{-r^2}r^2dr\big) ^2\\
 &\hspace*{1.2in}+2\big( \Lambda e^{-\Lambda ^2}+\int _{\Lambda }^{+\infty }e^{-r^2}dr\big)\big( \Lambda ^5e^{-\Lambda ^2}+5\int _{\Lambda }^{+\infty }e^{-r^2}r^4dr\big) \Big) \\
 = & \hspace*{0.1in}e^{-\Lambda ^2}\frac{2-e^{-\Lambda ^2}}{(1-e^{-\Lambda ^2})^2}\Big( \big( (\Lambda ^3+\frac{3}{2}\Lambda )e^{-\Lambda ^2}+\frac{3}{2}\int _{ \Lambda }^{+\infty }e^{-r^2}dr\big) ^2\\
 &\hspace*{1.2in}+2\big( \Lambda e^{-\Lambda ^2}+\int _{\Lambda }^{+\infty }e^{-r^2}dr\big)\big( (\Lambda ^5+\frac{5}{2}\Lambda ^3+\frac{15}{4}\Lambda )e^{-\Lambda ^2}+5\int _{\Lambda }^{+\infty }e^{-r^2}dr\big) \Big) \\
 \leq & \hspace*{0.1in}e^{-3\Lambda ^2}\frac{2-e^{-\Lambda ^2}}{(1-e^{-\Lambda ^2})^2}\Big( \big( \Lambda ^3+\frac{3}{2}\Lambda +\frac{3}{8}\sqrt{\pi }\big) ^2+2\big( \Lambda +\frac{\sqrt{\pi}}{4}\big) \big( \Lambda ^5+\frac{5}{2}\Lambda ^3+\frac{15}{4}\Lambda )\Big) .
\end{align*}
Hence $c_1-c_x\alpha \kappa <0$ for $\Lambda $ large enough.\\
Consequently, (\ref{1st-difference})-(\ref{2nd-difference}) give separate estimates for $|h_{kx}|^2$ and $|h_{k2}|^2$, leading to
 \begin{align*}
 \int_0^T(|h_{kx}|^2+|h_{kx}|^2)(t)dt\leq c&\Big(
 \int \frac{P}{1+P}|h_{k}(T,p)|^2dp+\int \frac{P}{1+P}|h_{k}(0,p)|^2dp\\
 &+\int _0^T\int \frac{P}{1+P}|h_{k\perp }(t,p)|^2dpdt+\gamma ^4(|G_{kx}|^2+|G_{k2}|^2+|G_{k1}|^2)
\Big) ,\quad k\neq 0.
 \end{align*}
Finally for $k= 0$, $\partial_t h_{02}= g\gamma ^2G_{02}$, $\partial_t h_{0x}= \gamma ^2G_{0x}$, so that by (\ref{hypG})-(\ref{hyp-h0}),
\begin{eqnarray*}
|h_{02}|^2+|h_{0x}|^2= 0,\quad t\in (0,T).
\end{eqnarray*}
We conclude that
\begin{align*}
\int _0^T\parallel h_\parallel (t,\cdot)\parallel _{2,2}^2dt\leq c\Big( \parallel h(T,\cdot)\parallel _{2,2}^2+\parallel h_0\parallel _{2,2}^2+\parallel \nu ^{\frac{1}{2}}h_\perp \parallel _{T,2,2}^2+\gamma ^4\parallel \nu^{-\frac{1}{2}}G\parallel _{T,2,2}^2\Big) .
\end{align*}
Using Lemma 3.2 leads to the statement of the lemma.\cqfd
Moreover, it follows from the expression of $h_\parallel $ in the basis $\{ (1+P)p_x, (1+P)(\lvert p\rvert ^2+gn_0)\} $ of the kernel of $L$, that there is a constant $c$ such that for any $h\in L^2_{\frac{P}{1+P}}$,
\begin{equation}\label{s-parallel-nu-or-not}
\parallel \nu ^{\frac{1}{2}}h_\parallel \parallel _{2,2}\leq c\parallel h_\parallel \parallel _{2,2}.
\end{equation}
Consequently, lemmas \ref{control-h_perp}-\ref{control-h_parallel} give that
\begin{lemma}\label{control-h}  Under the assumptions (3.9) and (3.10), there is a constant $\tilde{\beta }_1>0$ such that the $x$-periodic solution $h$ to the initial value problem (\ref{h}) satisfies
\[ \begin{aligned}
&\parallel h(t,.)\parallel_{2,2}+\sqrt{\gamma }\parallel \nu^{\frac{1}{2}} h\parallel _{T,2,2}
\leq \tilde{\beta }_1\Big(  \parallel h_0\parallel_{2,2}+
\gamma ^{\frac{3}{2}} \parallel\nu^{-\frac{1}{2}}G_{\perp}\parallel_{T,2,2}+ \gamma\parallel G_{\parallel}\parallel_{T,2,2}\Big) ,\quad t\in[0,T].
\end{aligned}\]
\end{lemma}
\hspace{1cm}\\
Due to the periodic setting,  the integrals $\int_0^{2\pi}\partial_xGdx= \int_0^{2\pi}\partial_xh_0dx=0$, and so assumptions (\ref{0-g-fourier})-(\ref{0-in-fourier}) are satisfied by $\partial _xh$, solution to (\ref{eq-derivative-h}). Similarly to Lemma \ref{control-h}, it holds\\
\begin{lemma}\label{control-derivative-h}
Under the assumptions (3.9) and (3.10) and analogous ones for $\partial _xh_0$ and $\partial _xG$, there is a constant $\tilde{\beta }_2>0$ such that the $x$-periodic solution $h$ to the initial value problem (\ref{h}) satisfies
\[ \begin{aligned}
&\parallel \partial _xh(t,.)\parallel _{2,2}+\sqrt{\gamma }\parallel \nu^{\frac{1}{2}} \partial _xh\parallel _{T,2,2}\leq \tilde{\beta }_2\Big(  \parallel \partial _xh_0\parallel _{2,2}+
\gamma ^{\frac{3}{2}} \parallel\nu^{-\frac{1}{2}}\partial _xG_{\perp}\parallel _{T,2,2}+ \gamma\parallel \partial _xG_{\parallel}\parallel_{T,2,2}\Big), \hspace*{0.01in} t\in[0,T].
\end{aligned}\]
\end{lemma}
In the rest of the paper, the notation
\begin{equation}\label{beta1}
\beta _1= \max \{ \tilde{\beta }_1, \tilde{\beta _2}\}
\end{equation}
will be used.
\hspace*{0.2in}\\
%
%
%
%
%
%
%
%
%
\section{ Proof of the main theorem.}
\setcounter{equation}{0}
\setcounter{theorem}{0}
We shall now use the a priori estimates for the linearized equations in Section 3 to construct solutions to the two component model (1.6-9), and begin with local in time solutions $\tilde{\Phi }$ and $\tilde{R}$ to the equations
\begin{equation}\label{eq-Phitilde}
\partial_t\tilde{\Phi } -i\partial^2_x\tilde{\Phi }= S_1(\tilde{R})\tilde{\Phi }-ign_0\bar{\tilde{\Phi }}+U(\tilde{\Phi },\tilde{R}) ,\quad \tilde{\Phi } (0,\cdot )= \tilde{\Phi } _i,
\end{equation}
\begin{equation}\label{eq-Rtilde}
\partial_t\tilde{R}+p_x\partial_x\tilde{R}= g\gamma \Big( n_0 L\tilde{R}+\gamma \big( L_{1,\tilde{\Phi }}\tilde{R}+Q_{1,\tilde{\Phi }}(\tilde{R},\tilde{R})\big) \Big) ,\quad \tilde{R}(0,\cdot ,\cdot )= \tilde{R}_i,\quad
\end{equation}
obtained from (\ref{eq-psi}), and (\ref{eq-f}). Here,
\[ \begin{aligned}
&S_1(\tilde{R})= -ign_0-\frac{g}{2}\gamma ^2\int PL\tilde{R}dp-2ig\gamma \int P\tilde{R}dp-\frac{g}{2}{\gamma ^3}\int Q(\tilde{R},\tilde{R})dp,\\
\hspace{1cm}\\
&U(\tilde{\Phi },\tilde{R})= -\frac{\sqrt{n_0}}{2}g\gamma \int PL\tilde{R}dp-2ig\sqrt{n_0}\int P\tilde{R}dp-i\sqrt{n_0}g\gamma (2\lvert \tilde{\Phi }\rvert ^2+\tilde{\Phi }^2)\\
&\hspace*{0.7in}-\frac{\sqrt{n_0}}{2}g\gamma ^{2}\int Q(\tilde{R},\tilde{R})dp
-ig\gamma ^2\tilde{\Phi }\lvert \tilde{\Phi }\rvert ^2,\\
\hspace*{0.2in}\\
&L_{1,\tilde{\Phi } }\tilde{R}:= \Big( \sqrt{n_0}(\tilde{\Phi }+\bar{\tilde{\Phi }})+\gamma \lvert \tilde{\Phi }\rvert ^2\Big) Ls= \frac{|\psi|^2-n_0}{\gamma }L\tilde{R}=:L_{1,\psi}\tilde{R},\\
\hspace*{0.2in}\\
&Q_{1,\tilde{\Phi }}(\tilde{R},\tilde{R})= (n_0+\gamma \sqrt{n_0}(\tilde{\Phi }+\bar{\tilde{\Phi }})+\gamma ^2\lvert \tilde{\Phi } \rvert ^2)\frac{Q(\tilde{R},\tilde{R})}{P}.
\end{aligned}\]
Denote by
\[ \begin{aligned}
&\parallel \tilde{\Phi } \parallel _{\infty T}:= \sup _{t\in [ 0,T] }\parallel \tilde{\Phi }(t,.)\parallel _{H^1(0,2\pi)},\\
&\parallel \tilde{R}\parallel _{q, 2,H^1}:= \big( \int_0^T\parallel \tilde{R}(t,\cdot ,\cdot )\parallel ^q_{ L^2_{\frac{P}{1+P}}(\mathbb{R}^3;H^1(0,2\pi))}dt\big) ^{\frac{1}{q}},\quad q\in [ 2,+\infty ] ,\\
& \parallel \tilde{R}(t)\parallel _{2,H^1}:=  \parallel \tilde{R}(t,\cdot ,\cdot )\parallel _{ L^2_{\frac{P}{1+P}}(\mathbb{R}^3;H^1(0,2\pi))} .
\end{aligned}\]
Some constants to be used later, are introduced next. By rescaling it is enough to consider $n_0= 1$. The constant $\beta _1$ was defined in (\ref{beta1}). Set
\begin{equation}\label{betas2}
\beta _2= \Big( \int P(1+P)dp\Big) ^{\frac{1}{2}},\quad \beta _2^\prime = \Big( \int P(1+P)\nu dp\Big) ^{\frac{1}{2}}.
\end{equation}
Denote by $\beta _3$ (resp. $\beta _4$) the norm of the injection from $H^1(0,2\pi )$ into $L^4(0,2\pi )$ (resp. $L^\infty (0,2\pi )$). By the Cauchy-Schwartz inequality and Lemmas 2.2 to 2.6, there are constants $(\beta _i)_{5\leq i\leq 8}$ such that for any function $g\in L^2_{\frac{P}{1+P}}$ (resp. $g\in L^2_{\nu \frac{P}{1+P}}$, $h \in L^2_{\frac{P}{1+P}}$),
\begin{equation}\label{L-seul}
\lvert  \int PLgdp\rvert \leq \beta _5\big( \int g^2\frac{P}{1+P}dp \big) ^{\frac{1}{2}},
\end{equation}
(resp.
\begin{equation}\label{L-nu}
(\int  \nu ^{-1}\frac{P}{1+P}(Lg)^2dp)^{\frac{1}{2}} \leq \beta _6(\int \nu \frac{P}{1+P}(g_\perp )^2dp)^{\frac{1}{2}},
\end{equation}
\begin{equation}\label{Q-seul}
\mid \int Q(g,g)dp\mid \leq \beta _7\big( \int g^2\frac{P}{1+P}dp \big) ^{\frac{1}{2}}\big( \int \nu (p)g^2\frac{P}{1+P}dp \big) ^{\frac{1}{2}},
\end{equation}
\begin{equation}\label{Q-nu}
(\int \nu^{-1}\frac{P}{1+P}(\frac{Q(g,h)}{P})^2dp)^{\frac{1}{2}}\leq \beta _8 \Big( \int \nu \frac{P}{1+P}g^2(p)dp\int \frac{P}{1+P}h^2(p)dp\Big) ^{\frac{1}{2}}).
\end{equation}
A control of $\parallel \nu^{-\frac{1}{2}}\frac{Q(g,h)}{P}\parallel _{2,2,H^1}$ will also be needed and is given in the following lemma.
\begin{lemma}\label{Q2}
\hspace*{0.1in}\\
For cylindrically symmetric functions $g$ and $h$ in $L^2(0,T; L^2_{\nu \frac{P}{1+P}}(\R ^3;H^1(0,2\pi )))$,
\begin{equation}\label{Q-norm-22H1}
\parallel \nu^{-\frac{1}{2}}\frac{Q(g,h)}{P}\parallel _{2,2,H^1}\leq 2\beta _4\beta _8(\parallel g\parallel _{\infty ,2,H^1}\parallel \nu ^{\frac{1}{2}}h \parallel _{2,2,H^1}+\parallel h \parallel _{\infty ,2,H^1}\parallel \nu ^{\frac{1}{2}}g \parallel _{2,2,H^1}).
\end{equation}
\end{lemma}
\underline{Proof of Lemma \ref{Q2}}\hspace*{0.3in}\\
By definition of $\parallel \cdot \parallel _{2,2,H^1}$,
\[ \begin{aligned}
&\parallel \nu^{-\frac{1}{2}}\frac{Q(g,h)}{P}\parallel _{2,2,H^1}^2\\
&\leq \int \nu^{-1}\big (\frac{Q(g,h)}{P}\big) ^2\frac{P}{1+P}dpdxdt+2\int \nu^{-1}\big( \frac{Q(g,\partial _xh)}{P}\big) ^2\frac{P}{1+P}dpdxdt+2\int \nu^{-1}\big( \frac{Q(\partial _xg,h)}{P}\big) ^2\frac{P}{1+P}dpdxdt\\
&\leq 2\beta _8^2\Big( \int \big( \int g^2\frac{P}{1+P}dp\big)  \big( \int \nu (h^2+(\partial _xh)^2)\frac{P}{1+P}dp\big) dxdt\\
&\quad \quad +\int \big( \int h^2\frac{P}{1+P}dp\big)  \big( \int \nu (g^2+(\partial _xg)^2)\frac{P}{1+P}dp\big) dxdt\Big) .
\end{aligned}\]
Moreover,
\[ \begin{aligned}
&\int \big( \int g^2\frac{P}{1+P}dp\big)  \big( \int \nu (h^2+(\partial _xh)^2)\frac{P}{1+P}dp\big) dxdt\\
&\leq \Big( \sup _{(t,x)\in [ 0,T]\times [ 0,2\pi ] }\int g^2(t,x,p)\frac{P}{1+P}dp\Big) \parallel \nu ^{\frac{1}{2}}h\parallel _{2,2,H^1}^2\\
&\leq \Big( \sup _{t\in [ 0,T]}\int (\sup _{x\in [ 0,2\pi ] }\lvert g(t,x,p)\rvert )^2\frac{P}{1+P}dp\Big) \parallel \nu ^{\frac{1}{2}}h\parallel _{2,2,H^1}^2\\
&\leq\beta _4^2 \Big( \sup _{t\in [ 0,T]}\int (g^2+(\partial _xg)^2)(t,x,p)\frac{P}{1+P}dxdp\Big) \parallel \nu ^{\frac{1}{2}}h\parallel _{2,2,H^1}^2\\
&= \beta _4^2 \parallel g\parallel _{\infty ,2,H^1}^2\parallel \nu ^{\frac{1}{2}}h\parallel _{2,2,H^1}^2.
\end{aligned}\]
Applying an analogous inequality to $\int \big( \int h^2\frac{P}{1+P}dp\big)  \big( \int \nu (g^2+(\partial _xg )^2)\frac{P}{1+P}dp$ leads to (\ref{Q-norm-22H1}).\\
\hspace*{0.1in}\\
\hspace*{0.1in}\\
%
\setcounter{proposition}{1}
\begin{proposition}\label{local-sol}
\hspace*{0.1in}\\
Let $(\tilde{\Phi }_i, \tilde{R}_i)\in H_{{\rm per}}^1(0,2\pi)\times L_{\frac{P}{1+P}}^2(\mathbb{R}^3;H^1_{{\rm per}}(0,2\pi))$, satisfy
\begin{eqnarray}\label{cond1-Ri}
\int \tilde{R}_i(x,p)p_x{P}dxdp= \int \tilde{R}_i(x,p)(|p|^2+gn_0){P}dxdp= 0,
\end{eqnarray}
and
\begin{equation}\label{cond2-Ri}
\int (|\psi_i|^2-1+\gamma \int_{\mathbb{R}^3}{P}\tilde{R}_idp)dx=0.
\end{equation}
There are $T_0$, $\gamma _0 \in \hspace*{0.03in} ] 0,1] $ and $c_1\geq 1$, such that for $\gamma \in ] 0,\gamma _0] $, there is a unique solution
\begin{equation*}
(\tilde{\Phi },\tilde{R})\in \mathcal{C}_b([ 0,T_0] ;H_{{\rm per}}^1(0,2\pi))\times \mathcal{C}_b([ 0,T_0] ;L^2_{\frac{P}{1+P}}(\R^3;H_{{\rm per}}^1(0,2\pi)))
\end{equation*}
 to (4.1-2), for which
\begin{eqnarray*}
\tilde{R} \in L^2_{\frac{(1+|p|)^3}{P(1+P)}}([0,T_0] \times\R^3;H_{{\rm per}}^1(0,2\pi))),
\end{eqnarray*}
and
\begin{equation*}\label{bounds-Phi-s}
\parallel \tilde{\Phi } \parallel _{\infty T}+\parallel \tilde{R}\parallel _{\infty ,2,H^1}+\sqrt{\gamma }\parallel\nu^{\frac{1}{2}}\tilde{R}\parallel_{2,2,H^1}\leq c_1(\parallel \tilde{\Phi }_i\parallel _{H^1}+\parallel \tilde{R}_i\parallel _{2,H^1}),
\end{equation*}
where
\begin{equation}\label{c1}
c_1= \max \{ 1, 2(\beta _1+2e^{12\pi g^2})\} .
\end{equation}
Given $\eta_0>0$, when $\parallel \tilde{\Phi }_i\parallel _{H^1}+\parallel \tilde{R}_i\parallel _{2,H^1}$ is bounded by $\eta_0$, then $T_0$ and $\gamma _0$ can be taken to depend only on $\eta_0$.
\end{proposition}
\underline{Proof of Proposition 4.2.}\\
Denote by $\eta _0>0$ a bound of $\parallel\tilde{\Phi }_i\parallel _{H^1}+\parallel \tilde{R}_i\parallel _{2,H^1}$. Let $\bar{c}_1$ be the maximum of
\begin{equation}\label{c1bar-a}
30g(1+2\eta _0e^{12\pi g^2})(\beta _2+\beta _5+\beta _1\beta _4\beta _7\eta _0+2\eta _0e^{12\pi g^2})e^{60g^2(1+\beta _1^2\eta _0^2(\beta _2^2+\beta _5^2+(\beta _1\beta _4\beta _7\eta _0)^2))},
\end{equation}
\begin{equation}\label{c1bar-c}
1+4\beta _1^2\beta _4\beta _8(1+\beta _4)\eta _0(1+4\eta _0^2e^{24\pi g^2}),
\end{equation}
and
\begin{align}\label{c1bar-b}
12g\big( c_1\eta _0(3+c_1\eta _0+4\beta _2+\beta _5+4\beta _2^\prime \beta _4\beta _8(1+c_1\eta _0))+2\beta _2+\beta _5\big) e^{12g^2(3+c_1^2\eta _0^2((\beta _2+\beta _5)^2+(\beta _4\beta _7c_1\eta _0)^2))} \nonumber \\
\quad \quad +6\beta _1\beta _4c_1\eta _0\big( (\beta _6+4\beta _4\beta _8c_1\eta _0)(1+c_1\eta _0)+2\beta _4\beta _8(1+2c_1\eta _0+c_1^2\eta _0^2)+1\big) .
\end{align}
Let $\gamma _0$ (resp. $T_0$) be chosen so that
\begin{equation}\label{gamma-0}
\gamma _0\leq \min \{ 1,\frac{e^{-12\pi g^2}}{8\beta _1\beta _6\eta _0(1+\eta _0e^{12\pi g^2})}, \frac{e^{-24\pi g^2}}{(16\beta _1(1+\beta _4)\beta _6\eta _0(1+2\eta _0e^{12\pi g^2}))^2}, \frac{1}{4\beta _1\beta _6(1+2\beta _4)c_1\eta _0(2+c_1\eta _0)}, \frac{1}{16\bar{c}_1^2}  \} ,
\end{equation}
\begin{equation}\label{T-0}
T_0\leq 1,\quad T_0\leq \frac{1}{16\bar{c}_1^2}.
\end{equation}
Let us prove by induction the existence and uniqueness of sequences $(\tilde{\Phi }^n)$ in $C([0,T_0], H^1_{per}(0,2\pi )) $ and $(\tilde{R}^n)$ in $C([0,T_0],L^2_{\frac{P}{1+P}}(R^3,H^1_{per}(0,2\pi ))) $ solutions to
\begin{equation}
\tilde{\Phi }^0= 0,\quad \tilde{R}^0= 0,
\end{equation}
\begin{equation}\label{Phi-n+1}
\partial_t\tilde{\Phi } ^{n+1}-i\partial^2_x\tilde{\Phi } ^{n+1}= S_1(\tilde{R}^n)\tilde{\Phi } ^{n+1}-ig\overline{\tilde{\Phi }^{n+1}}+U (\tilde{\Phi } ^n,\tilde{R}^n),\quad \tilde{\Phi } ^{n+1}(0,x)= \tilde{\Phi } _i(x),
\end{equation}
\begin{equation}\label{s-n+1}
\partial_t\tilde{R}^{n+1}+p_x\partial_x\tilde{R}^{n+1}= g\gamma \Big( L\tilde{R}^{n+1}+\gamma \big( L_{1,\tilde{\Phi }^n}\tilde{R}^{n+1}+Q_{1,\tilde{\Phi }^n}(\tilde{R}^n,\tilde{R}^n)\big) \Big) ,\quad \tilde{R}^{n+1}(0,\cdot ,\cdot )= \tilde{R}_i,
\end{equation}
and such that for any $T\leq T_0$ and $\gamma \leq \gamma _0$,
\begin{align}\label{induc-deltaPhi-n}
\parallel \delta _n\tilde{\Phi } \parallel _{\infty T}+\parallel \delta _n\tilde{R}\parallel _{\infty ,2,H^1}
+\sqrt{\gamma }\parallel \nu^{\frac{1}{2}}\delta_n\tilde{R}\parallel_{2,2,H^1}\hspace*{2.9in}\nonumber \\
\hspace*{0.08in}\leq \bar{c}_1(\sqrt{T}+\sqrt{\gamma })(\parallel \delta _{n-1}\tilde{\Phi } \parallel _{\infty T}+\parallel \delta _{n-1}\tilde{R}\parallel _{\infty ,2,H^1}+\sqrt{\gamma }\parallel \nu^{\frac{1}{2}}\delta_{n-1}\tilde{R}\parallel_{2,2,H^1}),\quad
\end{align}
and
\begin{equation}\label{induc-Phi-n}
\parallel \tilde{\Phi } ^n\parallel _{\infty T}+\parallel \tilde{R}^n\parallel _{\infty ,2,H^1}+\sqrt{\gamma }\parallel \nu^{\frac{1}{2}}\tilde{R}^n\parallel_{2,2,H^1}\leq c_1(\parallel \tilde{\Phi }_i\parallel _{H^1}+\parallel \tilde{R}_i\parallel _{2,H^1}),\quad n\in \N .
\end{equation}
Here $\delta _n\tilde{\Phi } =  \tilde{\Phi } ^{n+1}-\tilde{\Phi } ^n$ and $\delta _n\tilde{R}= \tilde{R}^{n+1}-\tilde{R}^n$.\\
The existence and uniqueness of $\tilde{\Phi } ^1$ in $C([ 0,T_0] , H^1_{per}(0,2\pi ))$ follow from Lemma 3.1. By (3.2) and the first bound on $T_0$ in (\ref{T-0}),
\begin{equation}\label{Phi-1}
\parallel \tilde{\Phi } ^1\parallel _{\infty T}\leq  2\parallel \tilde{\Phi } _i\parallel _{H^1}e^{12\pi g^2},\quad T\in ] 0,T_0] .
\end{equation}
The existence and uniqueness of $\tilde{R}^1$ in $C([ 0,T_0] ,L^2_{\frac{P}{1+P}}(\R ^3, H^1_{per}(0,2\pi )))$) follow from analogous arguments for the solution of the linearized Boltzmann equation. Lemmas 3.4-5 apply to $\tilde{R}^1$ and $\partial_x \tilde{R}^1$, by the assumption (4.3) on $\tilde{R}_i$, and because by periodicity
\begin{eqnarray*}
\int \partial_x \tilde{R}^1(t,x,p)(|p|^2+g)Pdxdp=\int \partial_x \tilde{R}^1(t,x,p)p_xPdxdp=0.
\end{eqnarray*}
Hence,
\begin{equation}\label{order-s1}
\parallel \tilde{R}^1\parallel_{\infty ,2,H^1}+\sqrt{\gamma }\parallel \nu ^{\frac{1}{2}}\tilde{R}^1\parallel_{2,2,H^1}\leq \beta _1\parallel \tilde{R}_i\parallel_{2,H^1},
\end{equation}
so that
\begin{equation}\label{order-s1-b}
\parallel \tilde{\Phi }^1\parallel_{\infty ,T}+\parallel \tilde{R}^1\parallel_{\infty ,2,H^1}+\sqrt{\gamma }\parallel \nu ^{\frac{1}{2}}\tilde{R}^1\parallel_{2,2,H^1}\leq \frac{c_1}{2}\big( \parallel \tilde {\Phi }_i\parallel _{H^1}+\parallel \tilde{R}_i\parallel_{2,H^1}\big) .
\end{equation}
And so, (\ref{induc-Phi-n}) holds for $n= 1$.\\
The existence and uniqueness of $\tilde{\Phi } ^2$ in $C([ 0,T_0] , H^1_{per}(0,2\pi ))$ follow from Lemma 3.1, since $S_1(\tilde{R}^1)$ and $U(\tilde{\Phi } ^1, \tilde{R}^1)$ belong to $L^\infty (0,T_0; H^1_{per}(0,2\pi ))$, $H^1(0,2\pi )$ being an algebra.  By the proof of Lemma \ref{1st-control},
\[ \begin{aligned}
\parallel \delta _1\Phi \parallel _{\infty T}\leq 3\sqrt{T}\Big( \int _0^T\parallel \big( U(\Phi ^1,\tilde{R}^1)+(S_1(\tilde{R}^1)+ig)\Phi ^1\big) (r)\parallel _{H^1}^2dr\Big) ^{\frac{1}{2}}&e^{3(\int _0^{T}\parallel S^1(\tilde{R}^1)(r)\parallel _{H^1}^2dr+2\pi g^2)},\\
& T\in ] 0,T_0] .
\end{aligned}\]
Moreover, by (\ref{L-seul}) and (\ref{Q-seul}),
\[ \begin{aligned}
\parallel &U(\Phi ^1,\tilde{R}^1)(r)+(S^1(\tilde{R}^1)(r)+ig)\Phi ^1\parallel _{H^1}\\
&\leq 2g(1+\parallel \Phi ^1\parallel _{\infty T})\big( \beta _5+\beta _2+\beta _4\beta _7\gamma ^2\parallel \nu ^{\frac{1}{2}}\tilde{R}^1\parallel _{2,2,H^1}\big) \parallel \tilde{R}^1(r)\parallel _{2,H^1}\\
&+10g\parallel \Phi ^1\parallel _{\infty T}^2(1+\parallel \Phi ^1\parallel _{\infty T}),\quad r\in [ 0,1]  .
\end{aligned}\]
And so,
\begin{align}\label{delta1Phi}
&\parallel \delta _1\Phi \parallel _{\infty T}\nonumber \\
&\leq 30g\sqrt{T}(1+\parallel \Phi ^1\parallel _{\infty T})\Big( \big( \beta _5+\beta _2+\beta _4\beta _7\gamma ^2\parallel \nu ^{\frac{1}{2}}\tilde{R}^1\parallel _{2,2,H^1}\big) \parallel \tilde{R}^1\parallel _{\infty ,2,H^1}+\parallel \Phi ^1\parallel _{\infty T}^2\Big) \nonumber \\
&\hspace*{1.in}\times e^{60g^2(1+(\beta _2^2+\beta _5^2)\parallel \tilde{R}^1\parallel ^2_{2,2,H^1}+\beta _4^2\beta _7^2\gamma \parallel \tilde{R}^1\parallel ^2_{\infty,2,H^1}\parallel \nu ^{\frac{1}{2}}\tilde{R}^1\parallel ^2_{T,2,2})}\nonumber \\
&\leq 30g\sqrt{T}(1+2\eta _0e^{12\pi g^2})\Big( \big( \beta _2+\beta _5+\beta _1\beta _4\beta _7\eta _0\big) \parallel \tilde{R}^1\parallel _{\infty ,2,H^1}+2\eta _0e^{12\pi g^2}\parallel \Phi ^1\parallel _{\infty T}\Big) \nonumber \\
&\hspace*{1.in}\times e^{60g^2(1+\beta _1^2\eta _0^2(\beta _2^2+\beta _5^2+(\beta _1\beta _4\beta_7\eta _0)^2))}.
\end{align}
A solution $\tilde{R}^2\in C([ 0,T_0] ,L^2_{\frac{P}{1+P}}(\R ^3, H^1_{per}(0,2\pi )))$ to (\ref{s-n+1}) can be obtained as the limit of the sequence $(r^k)_{k\in \N}$ defined by
\begin{align}
&r^0= 0,\quad \partial _tr^{k+1}+p_x\partial _xr^{k+1}= g\gamma \Big( Lr^{k+1}+\gamma \big( L_{1\Phi ^1}r^k+Q_{1\Phi ^1}(\tilde{R}^1,\tilde{R}^1)\big) \Big) ,\nonumber \\
&r^k(0,\cdot ,\cdot )= \tilde{R}_i.\nonumber
\end{align}
By Lemma 3.4, (\ref{L-nu}) and the second condition in (\ref{gamma-0}),
\[ \begin{aligned}
\parallel \nu ^{\frac{1}{2}}\delta _kr\parallel _{T,2,2}&\leq \gamma \beta _1\beta _6(2+\parallel \Phi ^1\parallel _{\infty T})\parallel \Phi ^1\parallel _{\infty T}\parallel \nu ^{\frac{1}{2}}\delta _{k-1}r\parallel _{T,2,2}\\
&\leq \frac{1}{2}\parallel \nu ^{\frac{1}{2}}\delta _{k-1}r\parallel _{T,2,2}, \quad k\in \N ^*.
\end{aligned}\]
And so, by a contraction argument, the local existence and uniqueness of $\tilde{R}^2$ follow.  Applying Lemmas 3.4-5 to $\delta _1\tilde{R}$ leads to
\[ \begin{aligned}
&\parallel \delta _1\tilde{R}\parallel _{\infty ,2,H^1}+\sqrt{\gamma }\parallel \nu^{\frac{1}{2}}\delta_1\tilde{R}\parallel_{2,2,H^1}\\
&\leq \beta _1\gamma ^{\frac{3}{2}}\Big( \parallel (\Phi ^1+\bar{\Phi }^1+\gamma \lvert \Phi ^1\rvert ^2)\nu^{-\frac{1}{2}}L\delta_1\tilde{R}\parallel _{2,2,H^1}+\parallel (\Phi ^1+\bar{\Phi }^1+\gamma \lvert \Phi ^1\rvert ^2)\nu^{-\frac{1}{2}}L\tilde{R}^1\parallel _{2,2,H^1}\\
&\quad \quad \quad \quad +\parallel \nu ^{-\frac{1}{2}}Q_{1,\Phi ^1}(\tilde{R}^1,\tilde{R}^1)\parallel_{2,2,H^1}\Big) \\
&\leq \beta _1(1+\beta _4)\gamma ^{\frac{3}{2}}\Big(  2\beta _6\parallel \Phi ^1\parallel _{\infty T}(1+\parallel \Phi ^1\parallel _{\infty T})(\parallel \nu^{\frac{1}{2}}\delta_1\tilde{R}\parallel_{2,2,H^1}+\parallel \nu^{\frac{1}{2}}\tilde{R}^1\parallel_{2,2,H^1})\\
&\quad \quad \quad \quad +4\beta _4\beta _8(1+\parallel \Phi ^1\parallel ^2_{\infty T})\parallel \tilde{R}^1\parallel _{\infty ,2, H^1}\parallel \nu ^{\frac{1}{2}}\tilde{R}^1\parallel _{2,2,H^1}\Big) .
\end{aligned}\]
And so, by (\ref{Phi-1}), (\ref{order-s1}) and the third condition in (\ref{gamma-0}),
\begin{equation}\label{delta1s-a}
\parallel \delta _1\tilde{R}\parallel _{\infty ,2,H^1}+\frac{3}{4}\sqrt{\gamma }\parallel \nu^{\frac{1}{2}}\delta_1\tilde{R}\parallel_{2,2,H^1}\leq (\frac{1}{4}+4\beta _1^2\beta _4\beta_8(1+\beta _4)\eta _0(1+4\eta _0^2e^{24\pi g^2}))\gamma \parallel \nu ^{\frac{1}{2}}\tilde{R}^1\parallel _{2,2,H^1}) .
\end{equation}
It results from (\ref{delta1Phi}), (\ref{delta1s-a}) and the bounds from below (\ref{c1bar-a})-(\ref{c1bar-c}) on $\bar{c}_1$ that
\begin{align}\label{part-df-c2tilde}
&\parallel \delta _1\tilde{\Phi } \parallel _{\infty T}+\parallel \delta _1\tilde{R}\parallel _{\infty ,2,H^1}
+\sqrt{\gamma }\parallel \nu^{\frac{1}{2}}\delta_1\tilde{R}\parallel_{2,2,H^1}\nonumber \\
&\leq \bar{c}_1 (\sqrt{T}+\sqrt{\gamma })(\parallel \tilde{\Phi } ^1\parallel _{\infty ,T}+\parallel \tilde{R}^1\parallel _{\infty ,2,H^1}+\sqrt{\gamma }\parallel \nu^{\frac{1}{2}}\tilde{R}^1\parallel_{2,2,H^1}).
\end{align}
And so, (\ref{induc-deltaPhi-n}) holds for $n= 1$.\\
The existence and uniqueness of $(\tilde{\Phi } ^n)_{n\geq 3}$ in $C([ 0,T_0] , H^1_{per}(0,2\pi ))$ follow from Lemma 3.1, since by induction on $n$ and  $H^1(0,2\pi )$ being an algebra, $S_1(\tilde{R}^n)$ and $U(\tilde{\Phi } ^n, \tilde{R}^n)$ belong to $L^\infty (0,T_0; H^1_{per}(0,2\pi ))$. The existence and uniqueness of $(\tilde{R}^n)_{n\geq 3}$ in $C([ 0,T_0] ,L^2_{\frac{P}{1+P}}(\R ^3, H^1_{per}(0,2\pi )))$ follow from similar arguments to those previously used for the existence and uniqueness of $\tilde{R}^2$.
\\
Assuming (\ref{induc-deltaPhi-n})-(\ref{induc-Phi-n}) up to $n-1$, and using the second (resp. fourth) condition in (\ref{T-0}) (resp. (\ref{gamma-0})) {implies that (\ref{induc-Phi-n}) holds for $n$.}  Then $\delta _n\tilde{\Phi } $ and $\delta _n\tilde{R}$ satisfy
\[ \begin{aligned}
&\partial_t\delta _n\tilde{\Phi } -i\partial^2_x\delta _n\tilde{\Phi } = S_1(\tilde{R}^n)\delta _n\tilde{\Phi } -ig\overline{\delta _n \tilde{\Phi }}+\Gamma _{n-1},\quad \quad \quad \delta _n\tilde{\Phi} (0,x)= 0,\hspace*{0.2in}\\
&\partial_t\delta _n\tilde{R}+p_x\partial_x\delta _n\tilde{R}= g\gamma \Big( L\delta _n\tilde{R}+\gamma(L_{1,\tilde{\Phi }^n}\delta _n\tilde{R}+G_{n-1})\Big) ,\quad \delta _n\tilde{R}(0,x,p)= 0.
\end{aligned}\]
Here
\[ \begin{aligned}
\Gamma _{n-1}&= \tilde{\Phi } ^n\delta _{n-1}(S_1(\tilde{R}))+\delta _{n-1}U (\tilde{\Phi } ,\tilde{R}),\\
G_{n-1}&= \Big( \delta _{n-1}\tilde{\Phi } +\overline{\delta _{n-1}\tilde{\Phi } }+\gamma  (\tilde{\Phi }^n\overline{\delta _{n-1}\tilde{\Phi }}+\overline{\tilde{\Phi } ^{n-1}}\delta _{n-1}\tilde{\Phi } )\Big) L\tilde{R}^n+\delta _{n-1}Q_{1,\tilde{\Phi } }(\tilde{R},\tilde{R}),
\end{aligned}\]
 and $(G_{n-1})_\parallel = 0$. Applying Lemma 3.1, it holds that
 \begin{eqnarray*}
 \parallel \delta _n\tilde{\Phi } \parallel^2 _{\infty T}\leq 6T\Big( \int _0^T\parallel \Gamma _{n-1}(t)\parallel^2 _{H^1}dt\Big) e^{6T\int _0^T(\parallel S_1(\tilde{R}^n)(t,\cdot )\parallel _{H^1}^2+2\pi g^2)dt }.
 \end{eqnarray*}
 By definition of $S_1(\tilde{R})$ and $U(\tilde{\Phi } ,\tilde{R})$,
 \[ \begin{aligned}
\Gamma _{n-1}= &-\frac{g}{2}\gamma \tilde{\Phi } ^n \Big( \gamma \int PL\delta _{n-1}\tilde{R}dp+4i\int P\delta _{n-1}\tilde{R}dp+\gamma ^2\int (Q(\tilde{R}^n,\delta _{n-1}\tilde{R})+Q(\delta _{n-1}\tilde{R},\tilde{R}^{n-1}))dp\Big) \\
&-\frac{g}{2}\gamma \int PL\delta _{n-1}\tilde{R}dp-2ig\int P\delta _{n-1}\tilde{R}dp\\
&-ig\gamma \big( (2\bar {\tilde{\Phi }}^n+\tilde{\Phi } ^{n-1}+\tilde{\Phi } ^n)\delta _{n-1}\tilde{\Phi } +2\tilde{\Phi } ^{n-1}\delta _{n-1}\bar{\tilde{\Phi } }\big) \\
&-\frac{g}{2}\gamma ^2\int (Q(\tilde{R}^n,\delta _{n-1}\tilde{R})+Q(\delta _{n-1}\tilde{R},\tilde{R}^{n-1}))dp\\
&-ig\gamma ^2\big( (\tilde{\Phi } ^n+\tilde{\Phi } ^{n-1})\bar{\tilde{\Phi }}^n \delta _{n-1}\tilde{\Phi } +(\tilde{\Phi } ^{n-1})^2\delta _{n-1}\bar{\tilde{\Phi } }\big) .
 \end{aligned}\]
 It follows from the Cauchy-Schwartz inequality w.r.t. the $p$ variable and (\ref{Q-nu}) that for every \\
 $t\in [ 0,T] $,
 \[ \begin{aligned}
\Big( \int  (\int Q(\tilde{R}^n,\delta _{n-1}\tilde{R})(t,x,p)dp)^2dx\Big) ^{\frac{1}{2}}&\leq \beta _2^\prime (\int \frac{P}{1+P}\nu ^{-1}\Big( \frac{Q(\tilde{R}^n,\delta _{n-1}\tilde{R})}{P}\Big) ^2dpdx)^{\frac{1}{2}}\\
&\leq \beta _2^\prime \beta _8\Big( \sup _{x\in [ 0,2\pi ] }\parallel \tilde{R}^n(t,x,\cdot )\parallel _{L^2_{\frac{P}{1+P}}}\Big) \parallel \nu ^{\frac{1}{2}}\delta _{n-1}\tilde{R}(t)\parallel _{2,2}\\
&\leq \beta _2^\prime \beta _4\beta _8\parallel \tilde{R}^n(t)\parallel _{2,H^1}\parallel \nu ^{\frac{1}{2}}\delta _{n-1}\tilde{R}(t)\parallel _{2,2}.
 \end{aligned}\]
 Analogously, $\partial _x Q(\tilde{R}^n,\delta _{n-1}\tilde{R})$ being equal to $Q(\partial _x\tilde{R}^n,\delta _{n-1}\tilde{R})+Q(\tilde{R}^n,\partial _x\delta _{n-1}\tilde{R})$,
 \[ \begin{aligned}
&\Big( \int  (\partial _x\int Q(\tilde{R}^n,\delta _{n-1}\tilde{R})(t,x,p)dp)^2dx\Big) ^{\frac{1}{2}}\\
&\leq \beta _2^\prime \beta _4\beta _8\Big( \parallel \delta _{n-1}\tilde{R}(t)\parallel _{2,H^1}\parallel \nu ^{\frac{1}{2}}\partial _x\tilde{R}^n(t)\parallel _{2,2}+\parallel \tilde{R}^n(t)\parallel _{2,H^1}\parallel \nu ^{\frac{1}{2}}\partial _x\delta _{n-1}\tilde{R}(t)\parallel _{2,2}\Big) .
 \end{aligned}\]
 And so,
 \[ \begin{aligned}
 &\parallel \int Q(\tilde{R}^n,\delta _{n-1}\tilde{R})(t, \cdot ,p)dp\parallel _{H^1}\\
 &\leq 2\beta _2^\prime \beta _4\beta _8\Big( \parallel \tilde{R}^n(t)\parallel _{2,H^1}\parallel \nu ^{\frac{1}{2}}\delta _{n-1}\tilde{R}(t)\parallel _{2,H^1}+\parallel \nu ^{\frac{1}{2}}\tilde{R}^n(t)\parallel _{2,H^1}\parallel \delta _{n-1}\tilde{R}(t)\parallel _{2,H^1}\Big) .
 \end{aligned}\]
 Consequently,
 \[ \begin{aligned}
 \parallel \Gamma _{n-1}(t)\parallel _{H^1}
 \leq &\hspace*{0.02in}g\gamma \parallel \tilde{\Phi } ^n\parallel _{\infty T}\Big( (\beta _5\gamma +4\beta _2) \parallel \delta _{n-1}\tilde{R}(t)\parallel _{2,H^1}\\
 &\hspace{1.2in}+\beta _2^\prime\beta _4\beta _8 \gamma ^2\big( (\parallel \tilde{R}^{n-1}(t)\parallel _{2,H^1}+\parallel \tilde{R}^n(t)\parallel _{2,H^1})\parallel \nu ^{\frac{1}{2}}\delta _{n-1}\tilde{R}(t)\parallel _{2,H^1}\\
&\hspace*{1.2in}+( \parallel \nu ^{\frac{1}{2}}\tilde{R}^{n-1}(t)\parallel _{2,H^1}+\parallel \nu ^{\frac{1}{2}}\tilde{R}^n(t)\parallel _{2,H^1}) \parallel \delta _{n-1}\tilde{R}(t)\parallel _{2,H^1}\big) \Big) \\
&+g(\beta _5\gamma +2\beta _2)\parallel \delta _{n-1}\tilde{R}(t)\parallel _{2,H^1}\\
&+3g\gamma (\parallel \tilde{\Phi } ^{n-1}\parallel _{\infty T}+\parallel \tilde{\Phi } ^n\parallel _{\infty T})\parallel \delta _{n-1}\tilde{\Phi } (t)\parallel _{H^1}\\
&+g\beta _2^\prime \beta _4\beta _8\gamma ^2\big( (\parallel \tilde{R}^{n-1}(t)\parallel _{2,H^1}+\parallel \tilde{R}^n(t)\parallel _{2,H^1})\parallel \nu ^{\frac{1}{2}}\delta _{n-1}\tilde{R}(t)\parallel _{2,H^1}\\
&\hspace*{1.2in}+( \parallel \nu ^{\frac{1}{2}}\tilde{R}^{n-1}(t)\parallel _{2,H^1}+\parallel \nu ^{\frac{1}{2}}\tilde{R}^n(t)\parallel _{2,H^1}) \parallel \delta _{n-1}\tilde{R}(t)\parallel _{2,H^1}\big)  \\
&+2g\gamma ^2(\parallel \tilde{\Phi } ^{n-1}\parallel _{\infty T}^2+\parallel \tilde{\Phi } ^n\parallel _{\infty T}^2)\parallel \delta _{n-1}\tilde{\Phi } (t)\parallel _{H^1}.
 \end{aligned}\]
 Moreover, for any $t\in [ 0,T] $,
 \[ \begin{aligned}
 \parallel S_1(\tilde{R}^n)(t)\parallel _{H^1}\leq 2g\Big(  1+\gamma (\beta _5\gamma +\beta _2)\parallel \tilde{R}^n(t)\parallel _{2,H^1}+\beta _4\beta _7\gamma ^3\parallel \tilde{R}^n(t)\parallel _{2,H^1}\parallel \nu ^{\frac{1}{2}}\tilde{R}^n(t)\parallel _{2,H^1}\Big) .
 \end{aligned}\]
 And so, using (\ref{induc-Phi-n}) at steps $n-1$ and $n$, and the first condition in (\ref{T-0})
\begin{align}\label{bdd-delta-Phi-n}
 \parallel \delta _n\tilde{\Phi } \parallel _{\infty T}&\leq 6g\sqrt{T}e^{12g^2\big( 3+c_1^2\eta _0^2( (\beta _2+\beta _5)^2+(\beta _4\beta _7c_1\eta _0)^2) \big) } \nonumber \\
 &\Big( 2c_1\eta _0(3+2c_1\eta _0)\parallel \delta _{n-1}\tilde{\Phi } \parallel _{\infty T}+(c_1\eta _0(4\beta _2+\beta _5+2\beta _2^\prime \beta _4\beta _8(1+c_1\eta _0))+2\beta _2+\beta _5)\parallel \delta _{n-1}\tilde{R}\parallel _{\infty ,2,H^1} \nonumber \\
 &+2c_1\eta _0(1+c_1\eta _0)\beta _2^\prime \beta _4\beta _8\sqrt{\gamma }\parallel \nu^{\frac{1}{2}}\delta_{n-1}\tilde{R}\parallel_{2,2,H^1}\Big) .
 \end{align}
 Multiplying the equation for $\delta_n\tilde{R}$ by $(|p|^2+g)P$ (resp. $p_xP$) and integrating on $(0,2\pi )\times \mathbb{R}^3$, gives
 \[ \begin{aligned}
&\frac{d}{dt}\int \delta _n\tilde{R}(t,x,p)(|p|^2+g)Pdxdp= 0,\quad (resp. \hspace*{0.02in}&\frac{d}{dt}\int \delta _n\tilde{R}(t,x,p)p_xPdxdp= 0).
\end{aligned}\]
Indeed, it follows from Lemma 2.1 that
\begin{eqnarray*}
\int \big( L\delta _n\tilde{R}\big) (|p|^2+g)Pdp= \int (\delta _n\tilde{R})L\big( (|p|^2+g)(1+P)\big) \frac{P}{1+P}dp= 0.
\end{eqnarray*}
Similarly, the $\delta _{n-1}Q_{1\tilde{\Phi }}(\tilde{R},\tilde{R})(\lvert p\rvert ^2+g)P$ term vanishes after integration, by the $\delta _0$ factor in the definition of $Q(\tilde{R},\tilde{R})$.\\
Being zero initially, $\int \delta _n\tilde{R}(t,x,p)(|p|^2+g)Pdxdp$ and $\int \delta _n\tilde{R}(t,x,p)p_xPdxdp$ remain identically zero, so that Lemmas 3.4-5 apply to the equation for $\delta_n\tilde{R}$, and $\delta_n\partial_x\tilde{R}$. Hence,
\begin{equation}\label{inter-R}
\parallel \delta _n\tilde{R}\parallel_{\infty ,2,H^1}+\sqrt{\gamma }\parallel \nu ^{\frac{1}{2}}\delta _n\tilde{R}\parallel_{2,2,H^1}\leq \beta _1\gamma ^{\frac{3}{2}}\parallel \nu ^{-\frac{1}{2}}(L_{1,\Phi _n}\delta _ns+G_{n-1})\parallel _{2,2,H^1}.
\end{equation}
Moreover,
\[ \begin{aligned}
&\parallel \nu ^{-\frac{1}{2}}L_{1,\tilde{\Phi } ^n}\delta _n\tilde{R}\parallel _{2,2,H^1}\leq \beta _6(1+2\beta _4)\parallel \tilde{\Phi } ^n\parallel _{\infty T}(2+\parallel \tilde{\Phi } ^n\parallel _{\infty T})\parallel \nu ^{\frac{1}{2}}\delta _n\tilde{R}_\perp \parallel _{2,2,H^1}\\
&\hspace*{1.45in}\leq \beta _6(1+2\beta _4)c_1\eta _0(2+c_1\eta _0)\parallel \nu ^{\frac{1}{2}}\delta _n\tilde{R}_\perp \parallel _{2,2,H^1},\\
&\text{and, using (\ref{Q-norm-22H1}),}\hspace*{1.in}\\
&\parallel \nu ^{-\frac{1}{2}}G_{n-1}\parallel _{2,2,H^1}\\
&\leq 2\beta _4\Big( (2+\parallel \tilde{\Phi } ^{n-1}\parallel _{\infty T}+\parallel \tilde{\Phi } ^n\parallel _{\infty T})\parallel \delta _{n-1}\tilde{\Phi } \parallel _{\infty T}(\beta _6+4\gamma \beta _4\beta _8\parallel \tilde{R}^n\parallel _{\infty ,2,H^1})\parallel \nu ^{\frac{1}{2}}\tilde{R}^n\parallel_{2,2,H^1}\\
&\hspace*{0.6in}+2\beta _4\beta _8(1+2\parallel \tilde{\Phi } ^{n-1}\parallel _{\infty T}+\parallel \tilde{\Phi } ^{n-1}\parallel _{\infty T}^2)\big( (\parallel \nu ^{\frac{1}{2}}\tilde{R}^{n-1}\parallel _{2,2,H^1}+\parallel \nu ^{\frac{1}{2}}\tilde{R}^n\parallel _{2,2,H^1})\parallel \delta _{n-1}\tilde{R}\parallel _{\infty  ,2,H^1}\\
&\hspace*{2.6in}+(\parallel \tilde{R}^{n-1}\parallel _{\infty ,2,H^1}+\parallel \tilde{R}^n\parallel _{\infty ,2,H^1})\parallel \nu ^{\frac{1}{2}}\delta _{n-1}\tilde{R}\parallel _{2,2,H^1}\big) \Big) \\
&\leq 4\beta _4c_1\eta _0\gamma ^{-\frac{1}{2}}\Big( (\beta _6+4\beta _4\beta _8c_1\eta _0)(1+c_1\eta _0)\parallel \delta _{n-1}\tilde{\Phi } \parallel _{\infty T}\\
&\hspace*{0.6in}+2\beta _4\beta _8(1+2c_1\eta _0+c_1^2\eta _0^2)\parallel \delta _{n-1}\tilde{R}\parallel _{\infty ,2,H^1}+\sqrt{\gamma }\parallel\nu ^{\frac{1}{2}} \delta _{n-1}\tilde{R}\parallel _{2,2,H^1}\Big).
\end{aligned}\]
And so, using (\ref{bdd-delta-Phi-n}), the fourth condition on $\gamma _0$ in (\ref{gamma-0}), in order to move the
\begin{eqnarray*}
\beta _1\beta _6(1+2\beta _4)c_1\eta _0(2+c_1\eta _0)\gamma ^{\frac{3}{2}}\parallel \nu ^{\frac{1}{2}}\delta _n\tilde{R}_\perp \parallel _{2,2,H^1}
\end{eqnarray*}
 term from the r.h.s. of (\ref{inter-R}) to its l.h.s., and using the bound from below (\ref{c1bar-b}) of $\bar{c}_1$,
\begin{eqnarray}\label{df-c-tilde-2}
\parallel \delta _n\tilde{\Phi } \parallel _{\infty T}+\parallel \delta _n\tilde{R}\parallel _{\infty ,2,H^1}
+\sqrt{\gamma }\parallel \nu^{\frac{1}{2}}\delta_n\tilde{R}\parallel_{2,2,H^1}\hspace*{2.in}\nonumber \\
\leq \bar{c}_1(\sqrt{T}+\sqrt{\gamma })(\parallel \delta _{n-1}\tilde{\Phi } \parallel _{\infty T}+\parallel \delta _{n-1}\tilde{R}\parallel _{\infty ,2,H^1}+\sqrt{\gamma }\parallel \nu^{\frac{1}{2}}\delta_{n-1}\tilde{R}\parallel_{2,2,H^1}).
\end{eqnarray}
This proves the induction (\ref{induc-deltaPhi-n}) for $n$. \\
It follows from (\ref{induc-deltaPhi-n})-(\ref{induc-Phi-n}), and the fifth (resp. second) condition in (\ref{gamma-0}) (resp. (\ref{T-0})), that the sequence $(\tilde{\Phi } ^n, \tilde{R}^n)$ converges in \\
$ L^\infty (0,T_0 ;H_{{\rm per}}^1(0,2\pi))\times L^\infty (0,T_0 ;L^2_{\frac{P}{1+P}}(\R^3;H_{{\rm per}}^1(0,2\pi)))$ when $n\rightarrow +\infty $ to a solution
\begin{eqnarray*}
(\tilde{\Phi }, \tilde{R})\in L^\infty (0,T_0;H^1_{per}(0,2\pi ))\times L^\infty (0,T_0;L^2_{\frac{P}{1+P}}(\R ^3;H^1_{per}(0,2\pi )))
\end{eqnarray*}
 of (4.1-2), satisfying
\begin{eqnarray}\label{orders-Phi1-R1}
\parallel \tilde{\Phi }\parallel _{\infty T_0}+\parallel \tilde{R}\parallel _{\infty  ,2,H^1}+\sqrt{\gamma}\parallel \nu ^{\frac{1}{2}}\tilde{R} \parallel _{2 ,2,H^1}\leq c_1\big( \parallel \tilde{\Phi } _i\parallel _{H^1}+\parallel \tilde{R}_i\parallel _{2,H^1}\big) .\quad
\end{eqnarray}
The solution belongs to $W^{1,1}(0,T_0;H^1_{per}(0,2\pi ))\times W^{1,1}(0,T_0;L^2_{\frac{P}{1+P}}(\R ^3;H^1_{per}(0,2\pi )))$, hence
\begin{eqnarray*}
 (\tilde{\Phi } ,\tilde{R})\in C_b([ 0,T_0] ;H^1_{per}(0,2\pi ))\times C_b([ 0,T_0] ;L^2_{\frac{P}{1+P}}(\R ^3;H^1_{per}(0,2\pi )).
\end{eqnarray*}
The uniqueness of the solution to (4.1-2) follows similarly, considering the difference of two solutions.
\cqfd
%
%
%
%
The following lemma  on the kinetic and internal energies of $\psi$, will also be needed to prove the global in time existence result of Theorem 1.1.\\
\\
\setcounter{theorem}{2}
\begin{lemma}
The solution $(f,\psi )= (P(1+\gamma \tilde{R}), \psi)$ of (1.6-9) satisfies
\begin{align}\label{derivative-alpha}
&\frac{d}{dt}\int \Big( \lvert \partial _x\psi\rvert ^2+\frac{g}{2}(\lvert \psi\rvert ^2-1)^2\Big) (t,x)dx= 2ig\gamma \int (\bar{\psi}\partial _x\psi-\psi\partial _x\bar{\psi})\int P\partial _x\tilde{R}dpdx\quad \quad \quad \quad \quad \quad \quad \nonumber \\
&-\frac{g}{2}\gamma ^2\int (\bar{\psi}\partial _x\psi+\psi\partial _x\bar{\psi})\int (PL\partial _x\tilde{R}+\gamma \partial _xQ(\tilde{R},\tilde{R}))dpdx-g\gamma ^2 \int \rvert \partial _x\psi\rvert ^2 \int (PL\tilde{R}+\gamma Q(\tilde{R},\tilde{R}))dpdx\nonumber \\
&-g^2\gamma ^2\int \lvert \psi\rvert ^2(\lvert \psi\rvert ^2-1)\int (PL\tilde{R}+\gamma Q(\tilde{R},\tilde{R}))dpdx.\hspace{0.2in}
\end{align}
\end{lemma}
\underline{Proof of Lemma 4.3.}\\
Given (\ref{cons-mass}), equation (\ref{eq-psi}) satisfied by $\psi$ is
\begin{equation}\label{eq-psi-2}
\partial _t\psi-i\partial _x^2\psi= -\psi (\frac{D}{2}+iA),
\end{equation}
where
\begin{eqnarray*}
D= g\gamma ^2 \int (PL\tilde{R}+\gamma  Q(\tilde{R},\tilde{R}))dp, \quad A= g (\lvert \psi\rvert ^2-1+2\gamma  \int P\tilde{R}dp).
\end{eqnarray*}
Multiply $(\ref{eq-psi-2})$ (resp. the conjugate of $(\ref{eq-psi-2})$) by $\partial _t\bar{\psi }$ (resp. $-\partial _t\psi $ ), integrate on $[ 0,2\pi ] $ so that
\[ \begin{aligned}
\frac{d}{dt}\int \big( \lvert \partial _x\psi\rvert ^2+\frac{g}{2}(\lvert \psi\rvert ^2-1)^2\big) (t,x)dx= &\frac{i}{2}\int D(\psi\partial _t\bar{\psi}-\bar{\psi}\partial _t\psi)dx-2g\gamma \int (\psi\partial _t\bar{\psi}+\bar{\psi}\partial _t\psi)\int P\tilde{R}dpdx\\
= &2ig\gamma \int (\psi\partial ^2_x\bar{\psi}-\bar{\psi}\partial ^2_x\psi)\int P\tilde{R}dpdx+2g\gamma \int \lvert \psi\rvert ^2D\int P\tilde{R}dpdx\\
&+\frac{1}{2}\int (\psi\partial ^2_x\bar{\psi}+\bar{\psi}\partial ^2_x\psi)Ddx-\int \lvert \psi\rvert ^2ADdx\\
= &2ig\gamma \int (\bar{\psi}\partial _x\psi-\psi\partial _x\bar{\psi})\int P\partial _x\tilde{R}dpdx+2g\gamma \int \lvert \psi\rvert ^2D\int P\tilde{R}dpdx\\
&-\frac{1}{2}\int (\psi\partial _x\bar{\psi}+\bar{\psi}\partial _x\psi)\partial _xDdx-\int \lvert \partial _x\psi\rvert ^2Ddx-\int \lvert \psi\rvert ^2ADdx.
\end{aligned}\]
This proves the lemma. \cqfd
\hspace*{0.1in}\\
\hspace*{0.1in}\\
%
%
%
%
%
%
\underline{Proof of Theorem 1.1.}\\
Defining $s=e^{\zeta t}\tilde{R}$, we look for a solution $(\tilde{\Phi }, s)$ to the equations
\begin{equation}\label{eqn-Phi-s}
\partial_t\tilde{\Phi } -i\partial^2_x\tilde{\Phi }= S_1(s)\tilde{\Phi }-ig\bar{\tilde{\Phi }}+U(\tilde{\Phi },s) ,\quad \tilde{\Phi } (0,\cdot )= \tilde{\Phi } _i,
\end{equation}
\begin{equation}\label{eqn-s-Phi}
\partial_ts+p_x\partial_xs= g\gamma \Big( Ls+\gamma \big( L_{1,\tilde{\Phi }}s+Q_{1,\tilde{\Phi }}(s,s)\big) \Big) +\zeta s,\quad s(0,\cdot ,\cdot )= \tilde{R}_i,
\end{equation}
obtained from (\ref{eq-Phitilde}) and (\ref{eq-Rtilde}). \\
Here $\zeta $ will be the positive rate of an exponential in time decay of $\tilde{R}$. Set $c_\zeta = \frac{\zeta }{\gamma }$.\\
Let $\gamma _0$ be given as in Proposition 4.2 when the norm of the initial conditions is bounded by $\eta_0=1$.
Let $c_\zeta $ and $(c_i)_{2\leq i\leq 7}$ be the constants defined by
\begin{equation}\label{add-cond-zeta}
c_\zeta = \min \{ 1, \frac{10}{\sqrt{1+g(1+\beta _3^4)}}, \frac{g\nu _0}{4\beta _1(1+\sqrt{\nu }_0)}\} ,
\end{equation}
\begin{equation}\label{c2}
c_2= 4\beta _1\big( 4\beta _1\beta _4c_4(\beta _6+\frac{\beta _8}{\sqrt{\nu _0}})+1\big) ,
\end{equation}
\begin{equation}\label{c3}
c_3= \sqrt{2}gc_4\Big( 2\sqrt{\beta _2}+\beta _4+\beta _5\Big) ,
\end{equation}
\begin{equation}\label{c4}
c_4= \beta _4\big( \mathcal{M}_0+1\big) ^{\frac{1}{2}},
\end{equation}
\begin{equation}\label{c5}
c_5= 4\beta _4(\frac{1}{g}+2c_4^2)+g+\frac{gc_4^2}{\sqrt{c_\zeta }}(\beta _5+\beta _4\beta _7),
\end{equation}
\begin{equation}\label{c6}
c_6= g\beta _5(3\beta _4+\sqrt{g}),
\end{equation}
\begin{equation}\label{c7}
c_7= \frac{1}{10}+g\beta _4\beta _7(3\beta _4+\sqrt{g}).
\end{equation}
Additionally, it is required that
\begin{equation}\label{df-eta-0}
\eta _0= \min \{ \hspace*{0.03in}1, \hspace*{0.03in}\frac{1}{4\beta _1(\beta _6c_5+2\beta _4\beta _8c_4^2)}, \hspace*{0.03in}\frac{1}{4\beta _1(\beta _6c_5+4\beta _4c_4(\beta _6+\beta _8c_4+\frac{\beta _8}{\sqrt{\nu _0}}))}\hspace*{0.03in}\} ,
\end{equation}
and
\begin{equation}\label{add-cond-gamma-0}
\gamma _0\leq  \min\{ 1,  \frac{1}{\eta _0\sqrt{2}}, \frac{1}{5c_6}, \frac{1}{25c_7^2}\} .
\end{equation}
Assume that
\begin{equation}\label{hyp-pf}
\parallel \tilde{\Phi } _i\parallel _{H^1}\leq \frac{\eta\zeta }{10c_1},\quad \parallel \tilde{R} _i\parallel _{2,H^1}\leq \frac{\eta\zeta}{10c_1(1+c_2)(1+c_3)},\quad \text{with }\eta \leq \eta _0.
\end{equation}
Set
\begin{equation*}
\alpha(t)=\int \big( \lvert \partial _x\psi\rvert ^2+\frac{g}{2}(\lvert \psi\rvert ^2-1)^2\big) (t,x)dx,
\end{equation*}
the semiclassical Hamiltonian of a strictly one-dimensional gas of bosons of mass $\frac{1}{2}$ and contact interaction strength $g$.\\
To start, there is by Proposition 4.2 a unique solution to problem (4.1-2) on a time interval $[0,T_0]$. From the computation,
\begin{eqnarray*}
\int (\lvert \psi _i(x)\rvert ^2-1)^2dx= \gamma ^2\int ((\tilde{\Phi } _i+\overline{\tilde{\Phi } _i})(x)+\gamma \lvert \tilde{\Phi }  _i\rvert ^2)^2dx\leq 2\gamma ^2\int (\lvert \tilde{\Phi }  _i(x)\rvert ^2+\lvert \tilde{\Phi } _i(x)\rvert ^4)dx,
\end{eqnarray*}
it holds that
\begin{eqnarray*}
\alpha (0)\leq (1+g(1+\beta _3^4))\gamma ^2\parallel \tilde{\Phi }  _i\parallel _{H^1}^2,
\end{eqnarray*}
hence
\begin{align}\label{bdd-alpha0}
\alpha (0)\leq \gamma ^2\eta ^2,
\end{align}
by (\ref{hyp-pf}) and the second bound in (\ref{add-cond-zeta}). Consider the set of times $t_1\leq T_0$ such that on $[ 0,t_1) $ the solution of (4.1-2) exists and satisfies
\begin{equation}\label{global-1}
\alpha (t)\leq 2\gamma ^2\eta ^2 \quad \text{and}\quad \parallel s(t)\parallel _{2,2}^2+\parallel \partial _xs(t)\parallel _{2,2}^2+\gamma (\parallel\nu^{\frac{1}{2}}s\parallel^2_{T_1,2,2}+\parallel \nu ^{\frac{1}{2}}\partial _xs\parallel^2 _{T_1,2,2})\leq 2 \eta ^2.
\end{equation}
This set is nonempty by continuity. Denote by $T_1$ its upper bound. We shall next prove that $T_1= T_0$ and improve the bounds (\ref{global-1}), which will allow the solution to be continued beyond $T_0$. That result is a main step in the proof of global existence, since it will imply that a solution will, as long as it can be continued, stay within the bounds of (\ref{global-1}).\\
On $[0,T_1] \times [ 0,2\pi ] $,
 \begin{equation}\label{bdd-psi}
\lvert \psi(t,x)\rvert \leq \beta _4\Big( \int (\lvert \psi(t,x)\rvert ^2+\lvert \partial _x\psi(t,x)\rvert ^2)dx\Big) ^{\frac{1}{2}}\leq \beta _4\Big(\mathcal{ M}_0+\alpha (t)\Big) ^{\frac{1}{2}}\leq \beta _4\Big(\mathcal{ M}_0+1\Big) ^{\frac{1}{2}}= c_4,
\end{equation}
by the bound on $\alpha (t)$ in (\ref{global-1}), the second bound in (\ref{add-cond-gamma-0}) and the definition of $c_4$. Moreover,
\[ \begin{aligned}
\lvert \hspace*{0.04in}\lvert \psi(t,x)\rvert ^2-1-\frac{1}{2\pi }\int (\lvert \psi(t,y)\rvert ^2-1)dy\rvert &\leq \beta _4\Big( \int \big( \lvert \psi(t,u)\rvert ^2-1-\frac{1}{2\pi }\int (\lvert \psi(t,y)\rvert ^2-1)dy\big) ^2du\\
&\quad \quad \quad +\int (\psi\partial _x\bar{\psi }+\bar{\psi }\partial _x\psi )^2(t,u)du\Big) \\
&\leq \beta _4\big( \int (\lvert \psi(t,u)\rvert ^2-1)^2+4c_4^2\lvert \partial _x\psi (t,u)\rvert  ^2)du\big)  \\
&\leq \beta _4(\frac{2}{g}+4c_4^2)\alpha (t)\\
&\leq 4\beta _4(\frac{1}{g}+2c_4^2)\gamma \eta ,\quad t\in [ 0,T_1] .
\end{aligned}\]
Multiplying (\ref{eq-psi}) (resp. the conjugate of (\ref{eq-psi})) by $\bar{\psi }$ (resp. $\psi $), integrating w.r.t. $x$ and adding the resulting equations leads to
\[ \begin{aligned}
\mid \frac{d}{dt}\int (\lvert \psi(t,x)\rvert ^2-1)dx\mid &= g\gamma ^2\mid \int \lvert \psi (t,x)\rvert ^2\int (PL\tilde{R} +\gamma Q(\tilde{R} ,\tilde{R} ))dpdx\mid \\
&\leq gc_4^2\sqrt{2\pi }\gamma ^2e^{-\zeta t}\parallel s(t)\parallel _{2,2}(\beta _5+\beta _4\beta _7\gamma e^{-\zeta t}\parallel \nu ^{\frac{1}{2}}s(t)\parallel _{2,2}).
\end{aligned} \]
And so, using the Cauchy-Schwartz inequality when integrating the previous inequality on $[ 0,t]$,
\[ \begin{aligned}
\lvert \hspace*{0.04in}\lvert \psi(t,x)\rvert ^2-1\rvert \leq 4\beta _4(\frac{1}{g}+2c_4^2)\gamma \eta &+\frac{1}{2\pi }\lvert \int \hspace*{0.04in}(\lvert \psi _i(x)\rvert ^2-1)dx\rvert +\frac{g}{\sqrt{2\pi }}c_4^2\beta _5\gamma ^2(2\zeta )^{-\frac{1}{2}}\parallel {s\parallel _{T_1,2,2}}\\
&+\frac{g}{\sqrt{2\pi }}c_4^2\beta _4\beta _7\gamma ^3(4\zeta )^{-\frac{1}{2}}\sup _{t\in [ 0,T_1] }\parallel s(t)\parallel _{2,2}\parallel \nu ^{\frac{1}{2}}{s}\parallel _{2,2,H^1}.
\end{aligned}\]
Using (\ref{bdd-alpha0}) and the $s$-part of (\ref{global-1}), leads to
\begin{equation}\label{bdd-psi-minus-n0}
\lvert \hspace*{0.04in}\lvert \psi(t,x)\rvert ^2-1\rvert \leq c_5\gamma \eta ,
\end{equation}
with $c_5$ defined in (\ref{c5}).\\
On $[0,T_1]$, consider the function $s$ which is a solution to
\[ \begin{aligned}
&\partial_ts+p_x\partial_xs= g\gamma \Big( Ls+\gamma \big( L_{1\psi}s+e^{-\zeta t}\lvert
\psi\rvert ^2\frac{Q(s,s)}{P}+\frac{\zeta }{g\gamma ^2}s\big) \Big) ,\\
&s(0,x,p)= \tilde{R} _i(x,p).
\end{aligned}\]
By Lemma 3.4, (\ref{L-nu}), (\ref{Q-nu}), (\ref{bdd-psi}), (\ref{bdd-psi-minus-n0}), the third bound on $c_\zeta$ in (\ref{add-cond-zeta}), and the $s$-part of (\ref{global-1}),
\[ \begin{aligned}
&\sup _{t\in [ 0,T_1] }\parallel s(t)\parallel _{2,2}+\frac{3}{4}\sqrt{\gamma }\parallel \nu ^{\frac{1}{2}}s\parallel _{T_1,2,2}\\
&< \beta _1\big( \parallel \tilde{R} _i\parallel _{2,2}+\beta _6c_5\gamma ^{\frac{3}{2}}\eta \parallel \nu ^{\frac{1}{2}}s_\perp \parallel _{T_1,2,2}
+\beta _4\beta _8c_4^2\gamma ^{\frac{3}{2}}\parallel s\parallel _{\infty ,2,H^1}\parallel \nu ^{\frac{1}{2}}s \parallel _{T_1,2,2}\big) \\
&\leq \beta _1\Big( \parallel \tilde{R} _i\parallel _{2,2}+(\beta _6c_5+2\beta _4\beta _8c_4^2)\gamma ^{\frac{3}{2}}\eta _0\parallel \nu ^{\frac{1}{2}}s\parallel _{T_1,2,2}\Big) .
\end{aligned}\]
If follows from the second bound on $\eta _0$ in (\ref{df-eta-0}) that
\begin{equation}\label{bdd-s}
\sup _{t\in [ 0,T_1] }\parallel s(t)\parallel _{2,2}+\frac{\sqrt{\gamma }}{2}\parallel \nu ^{\frac{1}{2}}s\parallel _{T_1,2,2}< \beta _1\parallel \tilde{R} _i\parallel _{2,2}.
\end{equation}
The function $\partial _xs$ satisfies
\[ \begin{aligned}
&\partial_t\partial _xs+p_x\partial_x\partial _xs= g\gamma \Big( L\partial _xs+(\lvert \psi(t,x)\rvert ^2-1)L\partial _xs+(\psi\partial _x\bar{\psi}+\bar{\psi}\partial _x\psi )Ls\\
& \hspace*{2.3in}+{e^{-\zeta t}}\partial _x(\lvert \psi\rvert ^2\frac{Q(s,s)}{P})+\frac{\zeta }{g\gamma ^2}\partial _xs\Big) ,\\
&\partial _xs(0,x,p)= \partial _x\tilde{R} _i(x,p).
\end{aligned}\]
Analogously, using Lemma 3.5,
\[ \begin{aligned}
\sup _{t\in [ 0,T_1) }\parallel \partial _xs(t)\parallel _{2,2}&+\frac{3}{4} \sqrt{\gamma }\parallel \nu ^{\frac{1}{2}}\partial _xs\parallel _{T_1,2,2}<\beta _1\Big( \parallel \partial _x\tilde{R} _i\parallel _{2,2}+\beta _6c_5\gamma ^{\frac{3}{2}}\eta \parallel \nu ^{\frac{1}{2}}\partial _xs\parallel _{T_1,2,2}\\
&+2c_4\sqrt{\gamma }\parallel (\partial _x\psi )\nu^{-\frac{1}{2}}Ls\parallel _{T_1,2,2}+\sqrt{\gamma }\parallel \nu^{-\frac{1}{2}}\partial _x(\lvert \psi\rvert ^2\frac{Q(s,s)}{P})\parallel _{T_1,2,2}\Big) .\\
\end{aligned}\]
Here,
\[ \begin{aligned}
\parallel (\partial _x\psi )\nu^{-\frac{1}{2}}Ls\parallel _{T_1,2,2}^2&\leq \beta _6^2\int _0^{T_1}\int \lvert \partial _x\psi\rvert ^2(t,x)\frac{P}{1+P}\nu (p)s^2(t,x,p)dpdxdt\\
&\leq \beta _6^2\Big( \sup _{t\in [ 0,T_1) }\int \lvert \partial _x\psi\rvert ^2(t,x)dx\Big) \Big( \int _0^{T_1}\int \frac{P}{1+P}\nu (p) \sup _{x\in [ 0,2\pi ] }s^2(t,x,p)dpdt\Big) \\
&\leq 2\beta _6^2\gamma ^2\eta ^2 \Big( \int _0^{T_1}\int \frac{P}{1+P}\nu (p) \sup _{x\in [ 0,2\pi ] }s^2(t,x,p)dpdt\Big) \quad \text{by   (\ref{global-1})}\\
&\leq 2\beta _4^2\beta _6^2\gamma ^2\eta ^2\Big( \parallel \nu ^{\frac{1}{2}}s\parallel _{T_1,2,2}^2+\parallel \nu ^{\frac{1}{2}}\partial _xs\parallel _{T_1,2,2}^2\Big) .
\end{aligned}\]
Moreover,
\[ \begin{aligned}
&\parallel \nu^{-\frac{1}{2}}\partial _x(\lvert \psi\rvert ^2\frac{Q(s,s)}{P})\parallel _{T_1,2,2}^2\\
&\leq 4c_4^2 \Big(\int _0^{T_1}\int \lvert \partial _x\psi\rvert ^2(t,x)\nu^{-1}(\frac{Q(s,s)}{P})^2(t,x,p)\frac{P}{1+P}dpdxdt\\
&\quad +c_4^2\int _0^{T_1}\int \nu^{- 1}(\frac{(Q(\partial_xs,s))}{P})^2(t,x,p)\frac{P}{1+P}dpdxdt\Big)\hspace{1in}\\
&\leq  4\beta _8^2c_4^2\Big( \frac{1}{\nu _0}\big( \sup _{t\in [ 0,T_1) }\int \lvert \partial _x\psi\rvert ^2(t,x)dx\big) \big( \int _0^{T_1}\int \nu (p) \sup _{x\in [ 0,2\pi ] }s^2(t,x,p)\frac{P}{1+P}dpdt\big)^2\hspace{1in} \\
&\quad +\beta _4^2c_4^2\parallel s \parallel^2 _{\infty,2,H^1}\parallel \nu^{\frac{1}{2}}\partial _xs\parallel _{T_1,2,2}^2 \Big) \\
&\leq 4\beta _4^2\beta _8^2c_4^2(\frac{2}{\nu _0}\gamma ^2\eta ^2\parallel \nu ^{\frac{1}{2}}s \parallel^4 _{2,2,H^1}+ c_4^2\parallel s \parallel^2 _{\infty,2,H^1}\parallel \nu^{\frac{1}{2}}\partial _xs\parallel _{T_1,2,2}^2)\\
&\leq 8\beta _4^2\beta _8^2c_4^2(\frac{2}{\nu _0}\gamma \eta ^4\parallel \nu ^{\frac{1}{2}}s \parallel^2 _{2,2,H^1}+ c_4^2\eta ^2\parallel \nu^{\frac{1}{2}}\partial _xs\parallel _{T_1,2,2}^2)
\end{aligned}\]
by (\ref{global-1}).
It follows from the third bound on $\eta _0$ in (\ref{df-eta-0}) that
\[ \begin{aligned}
\parallel \partial _xs(t)\parallel _{2,2}+\frac{\sqrt{\gamma }}{2}\parallel \nu ^{\frac{1}{2}}\partial _xs\parallel _{T_1,2,2}
\leq \beta _1\big( 8\beta _1\beta _4c_4(\beta _6+\frac{\beta _8}{\sqrt{\nu _0}})\parallel \tilde{R} _i\parallel _{2,2}+\parallel \partial _x\tilde{R} _i\parallel _{2,2}).
\end{aligned}\]
Consequently,
\begin{align}\label{global-2}
\parallel s(t)\parallel _{2,2}+\parallel \partial _xs(t)\parallel _{2,2}+\sqrt{\gamma }\big( \parallel \nu ^{\frac{1}{2}}s\parallel _{T_1,2,2}+\parallel \nu ^{\frac{1}{2}}\partial _xs\parallel _{T_1,2,2}\big) \nonumber\\
\leq c_2\big( \parallel \tilde{R} _i\parallel _{2,2}+\parallel \partial _x\tilde{R} _i\parallel _{2,2}\big) ,
\end{align}
with $c_2$ defined in (\ref{c2}). And so, using (\ref{hyp-pf}),
\begin{equation}\label{global-2b}
\parallel s(t)\parallel _{2,2}+\parallel \partial _xs(t)\parallel _{2,2}+\sqrt{\gamma }\big( \parallel \nu ^{\frac{1}{2}}s\parallel _{T_1,2,2}+\parallel \nu ^{\frac{1}{2}}\partial _xs\parallel _{T_1,2,2}\big)
\leq \frac{\eta\zeta}{10(1+c_3)},\hspace*{0.1in} \quad t\in [ 0,T_1] .
\end{equation}
{Recalling that $\zeta\leq 1$, this improves the second inequality in (\ref{global-1}).}\\
\\
Next consider $\alpha$. To improve the first inequality in (\ref{global-1}), each term in the right hand side of (\ref{derivative-alpha}) is first controlled separately. By (\ref{bdd-psi}) and the Cauchy-Schwartz inequality,
\begin{align}\label{df-c3}
\lvert 2ig\gamma &\int (\bar{\psi }\partial _x\psi -\psi\partial _x\bar{\psi })\int P\partial _x\tilde{R} dpdx-\frac{g}{2}\gamma ^2\int (\bar{\psi }\partial _x\psi +\psi\partial _x\bar{\psi })\int (PL\partial _x\tilde{R} +\gamma \partial_xQ(\tilde{R} ,\tilde{R} ))dpdx\rvert \quad \quad \quad \quad \quad \quad \nonumber  \\
&\leq gc_4\big( 2\sqrt{\beta _2}+\beta _5\gamma \big) \gamma e^{-\zeta t}\int \lvert \partial _x\psi(t,x)\rvert {(\int  \frac{ P}{1+P}}\lvert \partial _xs(t,x,p)\rvert^2dp)^{\frac{1}{2}}dx\hspace*{2.in}\nonumber \\
&\quad +2gc_4\beta _7\gamma ^3e^{-\zeta t}\int \lvert \partial _x\psi(t,x)\rvert (\int \frac{\nu P}{1+P}\lvert s(t,x,p)\rvert^2dp\int \frac{ P}{1+P}\lvert \partial _xs(t,x,p)\rvert^2dp)^{\frac{1}{2}}dx \nonumber \\
&\leq gc_4\gamma e^{-\zeta t}\sqrt{\alpha (t)}\parallel \partial _xs(t)\parallel _{2,2}\Big( 2\sqrt{\beta _2}+\beta _5+\gamma ^2e^{-\zeta t}\sup  _{x\in [ 0,2\pi ] }(\int \frac{\nu P}{1+P}s^2(t,x,p)dp)^{\frac{1}{2}}\Big) \hspace*{0.9in}\nonumber \\
&\leq gc_4\Big( 2\sqrt{\beta _2}+\beta _5+\beta _4\Big) \gamma e^{-\zeta t}\sqrt{\alpha (t)}\parallel \partial _xs(t)\parallel _{2,2} (1+\gamma (\parallel \nu^{\frac{1}{2}}s(t)\parallel _{2,2}
+\parallel \nu^{\frac{1}{2}}\partial _xs(t)\parallel _{2,2}))\nonumber\\
&= \frac{c_3}{{\sqrt{2}}}\gamma e^{-\zeta t}\sqrt{\alpha (t)}\parallel \partial _xs(t)\parallel _{2,2} (1+\gamma \parallel \nu^{\frac{1}{2}}s(t)\parallel _{2,H^1}),
\end{align}
by the definition (\ref{c3}) of $c_3$.
{And so, by the $\alpha $-part of (\ref{global-1}) and (\ref{global-2b}),
\begin{align}\label{global-3}
\lvert 2ig\gamma &\int (\bar{\psi }\partial _x\psi -\psi\partial _x\bar{\psi })\int P\partial _x\tilde{R} dpdx-\frac{g}{2}\gamma ^2\int (\bar{\psi }\partial _x\psi +\psi\partial _x\bar{\psi })\int (PL\partial _x\tilde{R} +\gamma \partial_xQ(\tilde{R} ,\tilde{R} ))dpdx\rvert \quad \quad \quad \quad \quad \quad \nonumber  \\
&\leq \zeta e^{-\zeta t}\frac{\gamma ^2\eta^2}{10}(1+\gamma \parallel \nu ^{\frac{1}{2}}s(t)\parallel _{2,H^1}).
\end{align}
}
By (\ref{L-seul}), the $\alpha $-part of (\ref{global-1}) and (\ref{global-2b}),
\begin{align}\label{global-4}
\mid \int \lvert \partial _x\psi \rvert ^2(t,x)\int PL\tilde{R} dpdx\mid &\leq \beta _5e^{-\zeta t}\alpha (t)\sup _{x\in [ 0,2\pi ] }(\int s^2(t,x,p)\frac{P}{1+P}dp)^{\frac{1}{2}}\nonumber \\
&\leq 2\beta _4\beta _5\gamma ^2\eta ^2e^{-\zeta t}\parallel s(t)\parallel _{{2,H^1}}\nonumber \\
&\leq \beta _4\beta _5\gamma ^2\eta ^3\zeta e^{-\zeta t},
\end{align}
and, by (\ref{Q-seul}), (\ref{global-2b}) and (\ref{global-1}),
\begin{align}\label{global-5}
&\mid \int \lvert \partial _x\psi \rvert ^2(t,x)\int Q(\tilde{R} ,\tilde{R} )dpdx\mid \nonumber\\
&\leq \beta _7e^{-2\zeta t}\alpha (t)\sup _{x\in [ 0,2\pi ] }(\int s^2(t,x,p)\frac{P}{1+P}dp)^{\frac{1}{2}}\sup _{x\in [ 0,2\pi ] }(\int \nu (p)s^2(t,x,p)\frac{P}{1+P}dp)^{\frac{1}{2}}\nonumber \\
&\leq \beta _4^2\beta _7e^{-2\zeta t}\alpha (t)\parallel s\parallel _{\infty ,2,H^1}\parallel \nu ^{\frac{1}{2}}s(t)\parallel _{2,H^1}\nonumber \\
&\leq \beta _4^2\beta _7\gamma ^2\eta ^3\zeta e^{-2\zeta t}\parallel \nu ^{\frac{1}{2}}s(t)\parallel _{2,H^1}.
\end{align}
The $\lvert \psi \rvert ^2(\lvert \psi \rvert ^2-1)$ factor of the integrand in the last term of the r.h.s. in (\ref{derivative-alpha}) is split into $\lvert \psi \rvert ^2(\lvert \psi \rvert ^2-1)= (\lvert \psi \rvert ^2-1)^2+(\lvert \psi \rvert ^2-1)$. It gives rise to the terms
\begin{eqnarray*}
\int (\lvert \psi \rvert ^2-1)^2\int (PL\tilde{R} +\gamma Q(\tilde{R} ,\tilde{R} ))dpdx \quad \text{and}\quad \int (\lvert \psi \rvert ^2-1)\int (PL\tilde{R} +\gamma Q(\tilde{R} ,\tilde{R} ))dpdx.
\end{eqnarray*}
Analogously to the previous control of $\int \lvert \partial _x\psi \rvert ^2\int (PL\tilde{R} +\gamma Q(\tilde{R} ,\tilde{R} ))dpdx$, it holds that
\[ \begin{aligned}
\mid \int (\lvert \psi \rvert ^2-1)^2\int (PL\tilde{R} +\gamma Q(\tilde{R} ,\tilde{R} ))dpdx\mid \leq \frac{2\beta _4}{g}\gamma ^2\eta ^3\zeta e^{-\zeta t}\big( \beta _5+\beta _4\beta _7\gamma e^{-\zeta t}\parallel \nu ^{\frac{1}{2}}s(t)\parallel _{2,H^1}\big) .
\end{aligned}\]
Moreover,
\[ \begin{aligned}
\mid &\int (\lvert \psi \rvert ^2-1)\int (PL\tilde{R} +\gamma Q(\tilde{R} ,\tilde{R} ))dpdx\mid \\
&\leq \sqrt{\alpha (t)}\Big( \int (\int (PL\tilde{R} +\gamma Q(\tilde{R} ,\tilde{R} ))dp)^2dx\Big) ^{\frac{1}{2}} \\
&\leq 2\sqrt{\frac{\alpha (t)}{g}}e^{-\zeta t}\big( \beta _5\parallel s(t)\parallel _{2,2}+\beta _4\beta _7\gamma e^{-\zeta t}\parallel s\parallel _{\infty ,2,H^1}\parallel \nu ^{\frac{1}{2}}s(t)\parallel _{2,2}\big) \\
&\leq \frac{1}{\sqrt{g}}\gamma \eta ^2\zeta e^{-\zeta t}\big( \beta _5+\beta _4\beta _7\gamma e^{-\zeta t}\parallel \nu ^{\frac{1}{2}}s(t)\parallel _{2,2}\big) .
\end{aligned}\]
{Consequently,
\begin{align}\label{global-6}
g^2\gamma ^2\lvert &\int \lvert \psi \rvert ^2(\lvert \psi \rvert ^2-1)\int (PL\tilde{R} +\gamma Q(\tilde{R} ,\tilde{R} ))dpdx\rvert \nonumber \\
&\leq g(2\beta _4+\sqrt{g})\gamma ^3\eta ^2\zeta e^{-\zeta t}\big( \beta _5+\beta _4\beta _7\gamma e^{-\zeta t}\parallel \nu ^{\frac{1}{2}}s(t)\parallel _{2,H^1}\big) .
\end{align}
And so, it follows from (\ref{derivative-alpha}), {(\ref{global-3}), (\ref{global-4}), (\ref{global-5}) and (\ref{global-6})} that
\begin{align}\label{bdd-alpha-prime}
\lvert \alpha ^\prime (t)\rvert &\leq (\frac{1}{10}+c_6\gamma )\gamma {^2}\eta ^2\zeta e^{-\zeta t} +c_7\gamma{^3}\eta ^2\zeta e^{-\zeta t} \parallel \nu ^{\frac{1}{2}}s(t)\parallel _{2,H^1},\quad t\in [ 0,T_1),
\end{align}
with $c_6$ (resp. $c_7$) defined in (\ref{c6}) (resp. (\ref{c7})). Integrating (\ref{bdd-alpha-prime}) on $[ 0,t] $ for $t\in [ 0,T_1)$, using the Cauchy-Schwartz inequality in the last term, (\ref{global-2b}) and the two last inequalities in (\ref{add-cond-gamma-0}) gives\\
\[ \begin{aligned}
\alpha(t)&\leq \alpha (0)+(\frac{1}{10}+c_6\gamma )\gamma ^2\eta ^2+c_7\gamma ^3\eta ^2\parallel \nu ^{\frac{1}{2}}s\parallel _{2,2,H^1}\\
&\leq \alpha (0)+\big( \frac{1}{10}+c_6\gamma +c_7\sqrt{\gamma }\eta \zeta \big) \gamma ^2\eta ^2\\
&\leq \alpha(0)+\frac{\gamma ^2\eta^2}{2},\quad t\in [ 0,T_1),
\end{aligned}\]
so that, by (\ref{bdd-alpha0}),
\begin{equation}\label{bdd-alpha}
\alpha(t)<\frac{3}{2}\gamma ^2\eta^2,\quad t\in [ 0,T_1).
\end{equation}
Since the bounds obtained in the r.h.s. of (\ref{global-2b}) and (\ref{bdd-alpha}) are better than the ones defining $T_1$ as the maximal time so that (\ref{global-1}) holds, it implies that $T_1= T_0$.\\
\hspace*{0.1in}\\
Let $T_2$ be the maximal time such that the solution exists on $[ 0,T_2[ $ and (\ref{global-1}) holds. \\
The family $(\tilde{\Phi }(t))_{t\in [ 0,T_2[ }$ is bounded in $H^1([ 0,2\pi ] )$. Indeed, the $\alpha $-part of (\ref{global-1}) implies that $\int _0^{2\pi }\lvert \partial _x\tilde{\Phi }(t,x)\rvert ^2dx\leq 2$. Moreover, (\ref{bdd-psi-minus-n0}) implies that
\begin{eqnarray*}
\gamma (\mathcal{R}\tilde{\Phi})^2+2\mathcal{R}\tilde{\Phi}+\gamma (\mathcal{I}\tilde{\Phi})^2-c_5\eta \leq 0,
\end{eqnarray*}
so that
\begin{eqnarray*}
\gamma \lvert \tilde{\Phi}(t,x)\rvert \leq 3+c_5,\quad (t,x)\in [ 0,T_2[ \times [ 0,2\pi ] .
\end{eqnarray*}
Moreover, it follows from (\ref{global-1}) that $(\tilde{R}(t))_{t\in [ 0,T_2[ }$ is bounded by $2$ in $H^1_{\frac{P}{1+P}}([ 0,2\pi ] \times \R ^3)$.\\
Consequently, Proposition 4.2 applies with any time $t<T_2$ as initial time and provides a unique solution to (\ref{eqn-Phi-s})-(\ref{eqn-s-Phi}) on an interval of time of length $\tilde{T}_0$ from Proposition 4.2 when the initial data are bounded by $\eta _0= 3(\frac{3+c_5}{\gamma }+1)$.
If $T_2$ is finite, using Proposition 4.2 with initial time $T_2-\frac{\tilde{T}_0}{2}$ and arguing as for (\ref{global-2b}), (\ref{bdd-alpha}), it follows that the solution can be continued beyond $T_2$ up to $T_2+\frac{\tilde{T}_0}{2}$, so that {(\ref{global-1}) holds. This contradicts $T_2$ being the maximal time. It results that $T_2= +\infty $.\\
\hspace*{0.1in}\\
Set $\lambda=\gamma^2$ and write {$\tilde{R}_i=\gamma R_i$, $\tilde{R}=\gamma R$, $\tilde{\Phi }_i=\gamma \Phi _i$, and $\tilde{\Phi }= \gamma \Phi $.} \\
The existence part of Theorem 1.1 is thus proved for $\lambda_1 $ in the statement of Theorem 1.1 given by  $\gamma _0^2$ with $\gamma _0$ defined in (\ref{gamma-0})-(\ref{add-cond-gamma-0}), $c_\zeta $ given by (\ref{add-cond-zeta}), and {$\eta_1=\frac{\eta }{10c_1(1+c_2)(1+c_3)} $ from} (\ref{hyp-pf}) with $\eta $ smaller than $\eta _0$ given by (\ref{df-eta-0}).
\hspace*{1.in}\\
It follows from (\ref{global-2}) that $R$ and $\partial_xR$ converge exponentially to zero
of order $\zeta$. As a consequence using the total mass conservation, $\int|\psi|^2dx$ converges exponentially to ${2\pi }n_0$. Using (\ref{bdd-alpha-prime}), it follows that $\alpha_{\infty}:=\lim_{t\rightarrow\infty} \alpha(t)$ exists, and is finite with the convergence to the limit being
exponential of order $\zeta $.
The solution $f$ is positive. Namely,  by (\ref{global-2b}) the magnitude of $R$ is bounded by $1$ in $L^{\infty}$, and so $|\lambda {R}|<1$.\cqfd
\newpage
\[\]
{\bf Bibliography}
\begin{enumerate}
\item \label{[A]}  Allemand, T.: Derivation of a two-fluids model for a Bose gas from a quantum kinetic system, Kin. Rel. Models 2, 379-402 (2009)
\item \label{[AN1]} Arkeryd, L., Nouri, A.: Bose condensates in interaction with excitations - a kinetic model, Comm. Math. Phys. 310, 765-788 (2012)
\item \label{[AN2]} Arkeryd, L., Nouri, A.: A Milne problem from a Bose condensate with excitations, Kin. Rel. Models 6, 671-686 (2013)
\item \label{[B]} Bose, S.N.: Planck's Law and Light Quantum Hypothesis, Z. Phys. 26, 178 (1924)
\item \label{[Eck]} Eckern, U.: Relaxation processes  in a condensate Bose gas, J. Low Temp. Phys. 54, 333-359 (1984)
\item \label{[E]} Einstein, A.: Sitzber. Kgl. Preuss. Akad. Wiss. 261 (1924)
\item \label{[EV]} Escobedo, M., Vel\'azques, J.: Finite time blow-up and condensation for the bosonic Nordheim equation, Invent. Math. (2013) doi:10.1007/s00222-014-0539-7
\item \label{[EPV]}Escobedo, M., Pezzotti, F., Valle, M.: Analytical approach to relaxation dynamics of condensed Bose gases, Ann. Phys. 326, 808-827 (2011)
\item \label{[GL]} Golse, F., Levermore, D.: Stokes-Fourier and acoustic limits for the Boltzmann equation; convergence proofs, Comm. Pure Appl. Math. 55, 336-393 (2002)
\item \label{[GNZ]} Griffin, A., Nikuni, T., Zaremba, E.: Bose-condensed gases at finite temperatures, Cambridge University Press, Cambridge 2009
\item \label{[HM]} Hohenberg, P., Martin, P.: Microscopic theory of superfluid helium, Ann. Phys. 34, 291-359 (1965)
\item \label{[IT]} Imamovic-Tomasovic, M., Kadanoff-Baym, L.: Kinetic theory for a trapped Bose condensate gas, Thesis, Univ. Toronto 2001
\item \label{[ITG]} Imamovic-Tomasovic, M., Griffin, A.: Quasiparticle kinetic equation in a trapped Bose gas at low temperature, J. Low Temp. Phys. 122, 617-655 (2001)
\item \label{[K]} Khalatnikov, I.M.: Theory of Superfluidity (in Russian), Nauka, Moskva (1971)
\item \label{[KD1]} Kirkpatrick, T.R., Dorfman, J.R.: Transport in a dilute but condensed nonideal Bose gas: kinetic equations, J. Low Temp. Phys. 58, 301-331 (1985)
\item \label{[KD2]} Kirkpatrick, T.R., Dorfman, J.R.: Transport coefficients in a dilute but condensed Bose gas, J. Low Temp. Phys. 58, 399-415 (1985)
\item \label{[KK]} Kane, J., Kadanoff, L.: Green's functions and superfluid hydrodynamics, Jour. math. Phys. 6, 1902-1912 (1965)
\item \label{[L]} Lu, X.: The Boltzmann equation for Bose-Einstein particles: condensation in finite time, J. Stat. Phys. 150, 1138-1176 (2013)
\item \label{[M]} Maslova, N.: Nonlinear evolution equations: kinetic approach, World Scientific, Singapore (1993)
\item \label{[N]} Nordheim, L.W.: On the kinetic methods in the new statistics and its applications in the electron theory of conductivity, Proc. Roy. Soc. London Ser. A 119, 689-698 (1928)
\item \label{[PS]} Pitaevski, L., Stringari, S.: Bose-Einstein Condensation, Clarendon Press, Oxford (2003)
\item \label{[PBMR]} Pomeau, Y., Brachet, M-\'E., M\'etens, S., Rica, S.: Th\'eorie cin\'etique d'un gaz de Bose dilu\'e avec condensat, CRAS 327 S\'erie II b, 791-798 (1999)
\item \label{[S]} Spohn, H.: Kinetics of the Bose-Einstein condensation, Physica D 239, 627-634 (2010)
\item \label{[ST]} Stoof, H.: Coherent versus incoherent dynamics during Bose-Einstein condensation in atomic gases, J. Low Temp. Phys. 114, 11-108 (1999)
\item \label{[UU]} Uehling, E.A., Uhlenbeck, G.E.: Transport phenomena in Einstein-Bose and Fermi-Dirac gases, Phys. Rev. 43, 552-561 (1933)
\item \label{[ZNG]} Zaremba, E., Nikuni, T., Griffin, A.: Dynamics of trapped Bose gases at finite temperatures, J. Low Temp. Phys. 116, 277-345 (1999)
\end{enumerate}
\end{document}